\documentclass[aip,jcp,amsmath,amssymb,longbibliography,reprint]{revtex4-1} 
\usepackage[T1]{fontenc}
\usepackage[latin9]{inputenc}
\usepackage{textcomp}
\usepackage{amsmath}
\usepackage{amssymb}
\usepackage{amsfonts}
\PassOptionsToPackage{normalem}{ulem}
\usepackage{ulem}
\usepackage{babel}
\usepackage{adjustbox}

\usepackage{graphicx}
\usepackage{float}

\usepackage{xcolor}
\usepackage{lineno}

\begin{document}

\title{Molecular conical intersections with odd electron number are realizations
of the topological Yang monopole}

\author{Chenchen Song}\email{ccsong@ucdavis.edu}
\affiliation{Department of Chemistry,   University of California Davis, Davis, CA 95616. }
\begin{abstract}
The two-level conical intersection of a molecule with odd electron
number in the absence of magnetic field obeys time-reversal symmetry
$T^{2}=-1$ (referred to as the $T^{2}=-1$ conical intersection),
and has a five-dimensional branching space due to the Kramer degeneracy.
Similar to how the conical intersection of a molecule in a magnetic
field ($T^{2}=0)$ behaves as the Dirac monopole, the $T^{2}=-1$
conical intersection behaves as the Yang monopole, an mathematical
generalization of Dirac monopole with SU(2) gauge field and SO(5)
symmetry. This implies that we can study topological properties of
$T^{2}=-1$ conical intersections in chemistry based on what is known
about Yang monopole in high energy physics. In this work, we present
a few mathematical tools to study this connection. First, we show
that geometric algebra and quaternion numbers together provides a
natural way to utilize the $T^{2}=-1$ time reversal symmetry and
the SO(5) symmetry in scaled coordinates, making it simple to derive
eigenfunctions, Berry connection and Berry curvature of $T^{2}=-1$
conical intersection. In particular, this approach provides a simple
proof that when viewed from the upper or lower states, the $T^{2}=-1$
conical intersection behaves as the self-dual or self-antidual Yang
monopole with second Chern number $C_{2}=+\frac{1}{4}$ or $-\frac{1}{4}$
respectively. In addition, we propose a visualization method for the
SU(2) Berry connection of $T^{2}=-1$ conical intersection. This is
achieved by showing that the nonzero part of the Berry connection
induces Hopf-fibration on the $S^{3}$ longitude space, and is further
visualized through stereographic projection.
\end{abstract}
\maketitle

\section{Introduction}

Conical intersection refers to a point where two electronic states
intersect and the degeneracy is lifted linearly in the displacements,
resulting in a characteristic double cone shapes of potential energy
surfaces. It has been discovered in a broad range of contexts, including
atomic physics, condensed matter physics, and quantum electrodynamics.\cite{larsonConicalIntersectionsPhysics2020}
In chemistry, we are interested in molecular conical intersections
that play an important role in photoreactions.\cite{yarkonyDiabolicalConicalIntersections1996,levineIsomerizationConicalIntersections2007a}
The simplest setup of molecular conical intersection is between two
adiabatic electronic states, and all the conical intersections throughout
this paper will refer to such two-level ones unless otherwise specified.
Based on time-reversal symmetry $T^{2}$, two-level conical intersections
can be classified into three types.\cite{domckeConicalIntersectionsElectronic2004}
When magnetic field is present, the molecule has no time reversal
symmetry $(T^{2}=0)$ and its conical intersection is of three-dimension.
Without magnetic field, molecules with even number of electrons obey
$T^{2}=+1$ and those with odd number of electrons obeys $T^{2}=-1$.
The corresponding conical intersections are two-dimensional and five-dimensional
respectively. It's worth noting that the aforementioned dimensions
assume the molecules have no other symmetry. For example, when $C_{s}$
point group symmetry is present, the branching space of $T^{2}=-1$
conical intersection is reduced to three dimension.\cite{matsikaSpinorbitCouplingConical2002}
Such point group symmetry of molecule will not be considered in this
work.

In this work, we are primarily interested in the geometric and topological
properties of the molecular conical intersections. The $T^{2}=+1$
conical intersection is the most studied one in chemistry with extensive
numerical simulations showing how it can affect excited state dynamics
through both topography\cite{pieriNonadiabaticNanoreactorAutomated2021a,virshupNonlinearDimensionalityReduction2012}
and topology \cite{curchodExactFactorizationPerspective2016,curchodDynamicsConicalIntersection2017}.
It gives rise to the molecular Aharonov-Bohm effect, and the wavefunction
changes sign after encircling the intersection point.\cite{aldenmeadMolecularAharonovBohm1980,meadDeterminationBornOppenheimer1979}
Recently, there has also been progress in the numerical simulations
of non-adiabatic properties for $T^{2}=0$ conical intersections.\cite{culpittNonadiabaticCouplingMatrix2024}
Because phase degree of freedom for $T^{2}=0$ case belongs to the
abelian U(1) group, the application of Stokes' Theorem leads to the
elegant result that the geometric phase along a closed loop is solely
determined by the enclosed surface area, giving rise to the famous
Berry's phase.\cite{berryQuantalPhaseFactors1984,zwanzigerBerryPhase1990}
In contrast to $T^{2}=0$ and $T^{2}=+1$ conical intersections that
are formed between two electronic states, the $T^{2}=-1$ conical
intersection involves four states. Due to Kramer degeneracy imposed
by $T^{2}=-1$ time-reversal symmetry,\cite{sakuraiModernQuantumMechanics2017a}
each energy level contains two degenerate wavefunctions that can be
mixed as the molecule moves in the configuration sp\textcolor{black}{ace,
resulting in a non-abelian SU(2) geometric phase,\cite{meadMolecularKramersDegeneracy1987}
also called the Wilczek-Zee} phase.\cite{wilczekAppearanceGaugeStructure1984}
Unlike its U(1) counterpart, there is no straightforward way of formulating
non-abelian Stokes' theorem \cite{gubarevNonAbelianStokesTheorem2004,kondoWilsonLoopMagnetic2008}
that is simple for computation. Due to these complications, our understanding
about the role of $T^{2}=-1$ molecular conical intersection in chemistry
is falling behind compared with the other two. In order to fill this
gap, we would like to take a step back and revisit the $T^{2}=-1$
conical intersection from a purely mathematical perspective.

The deep connection between geometric phase and the underlying topology
was first pointed out by Simon.\cite{simonHolonomyQuantumAdiabatic1983}
As summarized in Nakahara's textbook,\cite{nakaharaGeometryTopologyPhysics2003}
the geometry and topology of conical intersections are best understood
in the framework of principal fiber bundle $P(M,G)$. Here, $M$ is
the manifold of the parameter space, and $G$ is the gauge group of
the phase degree of freedom, such that a quantum state $\lvert\psi\rangle$
and $g\lvert\psi\rangle$, $\forall g\in G$ are equivalent. On the
other hand, as summarized by Yang,\cite{Yang1996} quantities in the
gauge field theory (e.g. electromagnetic theory, Yang-Mills theory)
can all be translated as concepts on the principal fiber bundle despite
the two being developed independently. Therefore, the three crucial
concepts for understanding conical intersection topology, i.e. Berry
connection (Eq.\ref{eq:Berry_connection}), Berry curvature (Eq.\ref{eq:curvature})
and geometric phase can be formally defined as the local connection
1-form, local curvature 2-form and holonomy on the principal bundle,
which are equivalent to gauge potential, gauge field and phase factor
in gauge theory. It's this one-to-one relation that allows theoretical
developments from gauge field theory in high energy physics to be
applied to the understanding of molecular conical intersections in
chemistry.

One well-known example to the above relation is that $T^{2}=0$ conical
intersection behaves topologically as Dirac monopole (i.e. magnetic
monopole)\cite{diracQuantisedSingularitiesElectromagnetic1997}. The
Dirac monopole is SO(3) symmetric, and its first Chern number can
take two possible quantized values $\pm g$ where $g$ depends on
the monopole strength.\cite{yangGeneralizationDiracMonopole1978}
The generalization of Dirac monopole is the SU(2) Yang monopole. The
latter is SO(5) symmetric, and its second Chern number can take two
possible quantized values $\pm a^{2}$ where $a$ depends on the monopole
strength.\cite{yangGeneralizationDiracMonopole1978} As pointed out
by Yang, three dimension is not sufficient for such monopole to exhibit
non-abelian nature, and five dimensions are needed to realize the
full complexity of the gauge group. Given the relation between $T^{2}=0$
conical intersection and Dirac monopole, one would naturally expect
that the $T^{2}=-1$ conical intersection behaves topologically as
the Yang monopole. In physics, this relation has been demonstrated
by considering the geometric phase of a spin-$3/2$ particle in a
quadrupole electric field.\cite{avronChernNumbersQuaternions1989,avronTopologicalInvariantsFermi1988}
More recently, Sugawa et.al reported an engineered Yang-monopole built
from conical intersection in the synthetic dimensions.\cite{sugawaSecondChernNumber2018,sugawaWilsonLoopWilczekZee2021}
The molecular conical intersections merely differ from these examples
in its parameter space being the molecular configuration space, but
are topologically the same. Therefore, the molecular conical intersection
with odd number of electrons in the absence of magnetic field is another
realization of Yang monopole. Compared with other examples, molecular
conical intersection is intrinsic to the molecule and does not involve
introducing external field or synthetic dimension\textcolor{black}{.
It's worth mentioning that molecular conical intersections are also
distinct in forming a high-dimensional ``seam'', although this aspect
is not studied in this work.}

The above discussions imply that in principle, we can understand the
topology of $T^{2}=-1$ molecular conical intersection from the properties
of Yang monopole in high energy physics. In particular, there should
exist a choice of gauge such that the corresponding Berry connection
and Berry curvature take the same form as the gauge potential and
field strength of Yang monopole derived in the original paper.\cite{yangGeneralizationDiracMonopole1978}
However, most of these works in theoretical physics emphasize more
on mathematical rigor and less on geometric intuition. Therefore,
the first goal of this work is to find a mathematical approach that
fully utilizes the symmetry and geometry of conical intersection,
such that its properties can be derived in a simple way. To achieve
this, we use a combination of geometric algebra and quaternion numbers.
The usage of geometric algebra (also called Clifford algebra) is inspired
by the following two properties. First, geometric algebra provides
a simple description of spatial rotation that is applicable to arbitrary
dimension,\cite{doranGeometricAlgebraPhysicists2003,lounestoCliffordAlgebrasSpinors2001}
making it a great tool to explore the rotational symmetry of the conical
intersection. Second, as explained in Section \ref{subsec:Geometric-algebra},
model Hamiltonian of conical intersections forms matrix representation
of a geometric algebra vector. This property is often used in topological
condensed matter physics by considering the extension of geometric
algebra, leading to the full classification of non-interacting topological
matters based on dimension and internal symmetry.\cite{chiuClassificationTopologicalQuantum2016,kitaevPeriodicTableTopological2009}
Our adoption of quaternion number is based on two considerations.
On one hand, systems with time reversal symmetry $T^{2}=+1,0,-1$
are known to be most naturally described by real, complex and quaternion
numbers respectively.\cite{baezDivisionAlgebrasQuantum2012} In particular,
because each quaternion number is mapped to a $2\times2$ complex
matrix, the $4\times4$ complex Hamiltonian matrix of $T^{2}=-1$
conical intersection can be compressed into a simpler $2\times2$
quaternion matrix, revealing the close resemblance between conical
intersections of different kinds. On the other hand, the isomorphism
of unit quaternion $\mathbb{H}^{1}\simeq S^{3}\simeq SU(2)$\cite{hsiangLecturesLieGroups2000}
allows us to translate naturally between geometry of the configuration
space and the associated SU(2) gauge group using the same quaternion
representation.

The second goal of this work is to develop a visualization method
such that the SU(2) Berry connection and Berry curvature of $T^{2}=-1$
conical intersection can be understood intuitively. Unlike $T^{2}=+1$
and $T^{2}=0$ cases that can be studied in the familiar $\mathbb{R}^{3}$
space, the five dimensional $T^{2}=-1$ conical intersections are
more challenging to envision. Fortunately, Yang's work suggests that
when the gauge is chosen properly, the nonzero part of the gauge potential
of Yang monopole can be fully understood on a $S^{3}$ longitude subspace.
Using stereographic projection, one can then visualize the four dimensional
$S^{3}$ by projecting it into $\mathbb{R}^{3}$. Stereographic projection
is commonly used in visualizing Hopf-fibration\cite{hopfUberAbbildungenDreidimensionalen1931}
$S^{1}\hookrightarrow S^{3}\to S^{2}$ due to its circle preserving
property, giving arise to the characteristic nested tori structure.\cite{lyonsElementaryIntroductionHopf2003}
We will use geometric algebra to showing that Berry connection in
fact induces a Hopf-fibration on the $S^{3}$-longitude space, thus
can be visualized in similar way as the Hopf fibration.

This paper is organized as follows. In Section \ref{sec:Math}, we
will start with briefly summarizing the maths used in this work, in
particular properties of geometric algebra and unit quaternion. These
ideas will be in Section \ref{sec:Conical-intersection} to provide
a simple way of solving the eigenfunctions for $T^{2}=-1$ conical
intersection. In Section \ref{sec:Topological}, we will study the
geometric and topological properties of $T^{2}=-1$ conical intersection,
in particular Berry connection and Berry curvature. We will first
solve its Berry connection using the geometric algebra approach in
Section \ref{subsec:Yang-monopole}, and show that they agree with
the gauge potential of Yang monopole. We then discuss how to visualize
the Berry connection using Hopf-fibration and stereographic projection
in Section \ref{subsec:Visualize-Berry-connection}. The properties
of Berry curvature and its visualization will be discussed in Section
\ref{subsec:Berry-curvature}.

\section{Mathematical background\label{sec:Math}}

In this section, we will give a brief review of the mathematical methods
used in this work, which are based on several textbooks.\cite{hsiangLecturesLieGroups2000,nakaharaGeometryTopologyPhysics2003,doranGeometricAlgebraPhysicists2003}
The matrix representations for several quantities in this section
are related to Pauli matrices, but there isn't a consistent convention
in literature. Therefore, we summarize the choice of matrix representations
used in this work in Table \ref{table:3Dvs5D}.

\begin{table*}[t]
\caption{Matrix representations of $Cl(3)$ vectors and bivectors,
unit quaternion $\mathbb{H}^{1}$, and $\mathfrak{su}(2)$ Lie algebra
through Pauli matrices. The right column shows the corresponding product, which
can be used to derive commutation and anti-commutation relations. }
\label{table:math}
{

\begin{tabular}{|c|c|c|}
\hline
 & Matrix representation & Product\tabularnewline
\hline
\hline
Pauli matrices & $\sigma_{1}=\left(\begin{array}{cc}
 & 1\\
1
\end{array}\right),\sigma_{2}=\left(\begin{array}{cc}
 & -i\\
i
\end{array}\right),\sigma_{3}=\left(\begin{array}{cc}
1\\
 & -1
\end{array}\right)$ & $\sigma_{i}\sigma_{j}=\delta_{ij}+i\epsilon_{ijk}\sigma_{k}$\tabularnewline
\hline
$Cl(3)$ vector & $\vec{e}_{i}=\sigma_{i}$ ( $\{x,y,z\}\leftrightarrow\{1,2,3\}$) & $\vec{e}_{i}\vec{e}_{j}=\delta_{ij}+\check{\beta}_{ij}$\tabularnewline
\hline
$Cl(3)$ Bivector  & $\check{\beta}^{k}=\epsilon_{ijk}\check{\beta}_{ij}=i\sigma_{k}$ & $\check{\beta}^{i}\check{\beta}^{j}=-\delta_{ij}-\epsilon_{ijk}\check{\beta}^{k}$\tabularnewline
 & $\check{\beta}_{yz}=\left(\begin{array}{cc}
 & i\\
i
\end{array}\right),\check{\beta}_{zx}=\left(\begin{array}{cc}
 & 1\\
-1
\end{array}\right),\check{\beta}_{xy}=\left(\begin{array}{cc}
i\\
 & -i
\end{array}\right)$ & \tabularnewline
\hline
Unit quaternion $\mathbb{H}^{1}$ & $\hat{q}_{i}=-i\sigma_{i}$ ($\{\hat{q}_{1},\hat{q}_{2},\hat{q}_{3}\}\leftrightarrow\{\hat{i},\hat{j},\hat{k}\}$) & $\hat{q}_{i}\hat{q}_{j}=-\delta_{ij}+\epsilon_{ijk}\hat{q}_{k}$\tabularnewline
 & $\hat{i}=\left(\begin{array}{cc}
 & -i\\
-i
\end{array}\right),\hat{j}=\left(\begin{array}{cc}
 & -1\\
1
\end{array}\right),\hat{k}=\left(\begin{array}{cc}
-i\\
 & i
\end{array}\right)$ & \tabularnewline
\hline
Lie algebra $\mathfrak{su}(2)$ & $\begin{array}{c}
\mathfrak{g}_{i}=-i\sigma_{i}\\
\mathfrak{g}_{1}=\hat{i},\mathfrak{g}_{2}=\hat{j},\mathfrak{g}_{3}=\hat{k}
\end{array}$ & $\mathfrak{g}_{i}\mathfrak{g}_{j}=-\delta_{ij}+\epsilon_{ijk}\mathfrak{g}_{k}$\tabularnewline
\hline
\end{tabular}
}
\end{table*}


\subsection{Geometric algebra\label{subsec:Geometric-algebra}}

This work will only use geometric algebra of N-dimensional Euclidean
space, which is denoted as $Cl(N)$. The geometric algebra is most
easily understood through its generators $\{\vec{e}_{i},1\leq i\leq N\}$
that anti-commute:
\begin{equation}
\{\vec{e}_{i},\vec{e}_{j}\}=2\delta_{ij}\label{eq:clifford}
\end{equation}
Vectors in geometric algebra will be denoted as $\vec{v}$, and can
be written as linear combination of the generators, i.e. $\vec{v}=\sum_{i}v_{i}\vec{e}_{i}$.
Eq.\ref{eq:clifford} involves taking products of geometric algebra
vectors, which is a combination of inner product (symmetric) and wedge
product (anti-symmetric)
\begin{equation}
\vec{u}\vec{v}=\vec{u}\cdot\vec{v}+\vec{u}\wedge\vec{v}
\end{equation}
In particular, products of orthogonal generators are $\vec{e}_{i}\vec{e}_{j}=\vec{e}_{i}\wedge\vec{e}_{j}$,
$i\neq j$. These form the basis of ``bivectors'', denoted as
\begin{equation}
\check{\beta}_{ij}=\vec{e}_{i}\wedge\vec{e}_{j}=\vec{e}_{i}\vec{e}_{j}\label{eq:bivector}
\end{equation}
Intuitively, $\check{\beta}_{ij}$ can be considered as the basis
of a directional surface element, which has unit area and points in
the positive direction of the plane spanned by $\vec{e}_{i}$ and
$\vec{e}_{j}$. The fact that the bivector of a plane can be formed
by taking product of two orthogonal unit vectors in that plane is
repeatedly used in this work.

As mentioned in the introduction, one of the major advantages of geometric
algebra is its simple description of rotations that can be generalized
to arbitrary high dimension. Instead of defining rotations through
rotational axis (which is unique to 3D), in high dimension it's more
appropriate to define rotation through rotating angle around the origin
within a plane while keeping all directions perpendicular to the plane
fixed. The ``rotor'' that describes rotating within the plane (identified
with bivector $\check{\beta}$) by angle $\theta$ is defined as
\begin{equation}
R_{\check{\beta}}(\theta)=e^{-\check{\beta}\frac{\theta}{2}}=\cos\frac{\theta}{2}-\sin\frac{\theta}{2}\check{\beta}\label{eq:rotor}
\end{equation}
A vector in geometric algebra then transforms under rotation as
\begin{equation}
R_{\check{\beta}}(\theta)\vec{v}R_{\check{\beta}}(-\theta)=e^{-\frac{\theta}{2}\check{\beta}}\vec{v}e^{\frac{\theta}{2}\check{\beta}}\label{eq:vector_rotation}
\end{equation}
On the other hand, if a quantity transforms under rotation as
\begin{equation}
R_{\check{\beta}}(\theta)\psi=e^{-\frac{\theta}{2}\check{\beta}}\psi\label{eq:spinor_rotation}
\end{equation}
then it is a ``spinor'' in geometric algebra. This property will
be used in Section \ref{subsec:Solve-modelH}, where Hamiltonian transforms
as vectors, while wavefunction transforms as spinor.

In the specific case of 3D space, the Pauli matrices in Table \ref{table:math}
satisfy the defining property of geometric algebra generators in Eq.\ref{eq:clifford},
thus they directly form the matrix representations of $Cl$(3) vector
generators. In addition, by comparing $\sigma_{i}\sigma_{j}$ with
$\vec{e}_{i}\vec{e}_{j}$, one can see that $\check{\beta}_{ij}=i\epsilon_{ijk}\sigma_{k}$
provides the matrix representation of the bivector basis. These will
be used for $T^{2}=0$ conical intersection in Section \ref{subsec:solve_3D}.
The fact that vector and bivector have similar matrix representations
is a unique property for $Cl(3)$, because they form duality in three-dimension,
i.e. every 1D line has a complement 2D plane.

\subsection{Unit quaternion $\mathbb{H}^{1}$ and 3-sphere $S^{3}$\label{subsec:quaternion}}

A quaternion number, denoted in this work with hat as $\hat{q}$,
can be considered as a generalization of complex number. It introduces
three imaginary numbers $\hat{i},\hat{j},\hat{k}$ with the following
relations:
\begin{equation}
\begin{array}{c}
\hat{i}^{2}=\hat{j}^{2}=\hat{k}^{2}=-1\\
\hat{i}\hat{j}=-\hat{j}\hat{i}=\hat{k},\hat{j}\hat{k}=-\hat{k}\hat{j}=\hat{i},\hat{k}\hat{i}=-\hat{i}\hat{k}=\hat{j}
\end{array}\label{eq:quaternion}
\end{equation}
A quaternion number can thus be written as $\hat{q}=r+a\hat{i}+b\hat{j}+c\hat{k}$.
If define $\hat{q}_{1}=\hat{i},\hat{q}_{2}=\hat{j},\hat{q}_{3}=\hat{k}$,
the above relations can be combined into
\begin{equation}
\hat{q}_{i}\hat{q}_{j}=-\delta_{ij}+\epsilon_{ijk}\hat{q}_{k}\label{eq:quaternion_commutation}
\end{equation}
In this work, we will primarily be using unit quaternion (denoted
as $\mathbb{H}^{1}$), where the coefficients satisfy $r^{2}+a^{2}+b^{2}+c^{2}=1$.
The coefficients $(r,a,b,c)$ thus can be considered as coordinates
of points on $S^{3}$ sphere. By identifying $1,\hat{i},\hat{j},\hat{k}$
with four orthogonal axes in $S^{3}\subset\mathbb{R}^{4}$, this provides
a one-to-one mapping between unit quaternion and $S^{3}$.

A quaternion with zero real part (i.e. $r=0$) is called a ``pure''
quaternion. An important property of pure unit quaternion is that
it behaves like a complex number, i.e.
\begin{equation}
\hat{n}^{2}=-1\text{ \ensuremath{\text{if}} }\hat{n}=n_{1}\hat{i}+n_{2}\hat{j}+n_{3}\hat{k},\text{\text{ }}n_{1}^{2}+n_{2}^{2}+n_{3}^{2}=1\label{eq:unit_pure_q}
\end{equation}
This allows one to define the exponential of $\hat{n}$ as follows,
which will be repeatedly used in this work
\begin{equation}
e^{\hat{n}\phi}=\cos\phi+\hat{n}\sin\phi\label{eq:exp_q}
\end{equation}

Start with the Pauli matrices in Table \ref{table:math}, one can
see that $q_{i}=-i\sigma_{i}$ satisfy the property of quaternion
imaginary numbers Eq.\ref{eq:quaternion_commutation}, thus forming
the corresponding complex matrix representations. This shows that
each quaternion number can be translated to a $2\times2$ complex
matrix. For example, the matrix representation of $e^{\hat{n}\phi}$
in Eq.\ref{eq:exp_q} can be written as
\begin{equation}
e^{\hat{n}\phi}=\left(\begin{array}{cc}
\cos\phi-i\sin\phi n_{3} & -\sin\phi n_{2}-i\sin\phi n_{1}\\
\sin\phi n_{2}-i\sin\phi n_{1} & \cos\phi+i\sin\phi n_{3}
\end{array}\right)\label{eq:exp_q_matrix}
\end{equation}
This will be used in Section \ref{subsec:solve_5D} to simplify the
Hamiltonian and wavefunction of $T^{2}=-1$ conical intersection.

It's worth mentioning that historically, quaternion was invented by
Hamilton to describe rotations in 3D.\cite{hankinsSirWilliamRowan1980}
The representations of rotations through quaternion and geometric
algebra are closely related but not identical (there is no differentiation
between vectors and bivectors in quaternion). To avoid confusion,
we will stick with geometric algebra representation for all rotations
including 3D.

\subsection{Lie group SU(2) and its Lie algebra $\mathfrak{su}(2)$ \label{subsec:SU(2)} }

As mentioned in the introduction, $T^{2}=-1$ conical intersection
has SU(2) gauge freedom due to Kramer degeneracy. Correspondingly,
the Berry connection and Berry curvature take value from its Lie algebra
$\mathfrak{su}(2)$. Geometrically, Lie algebra is the tangent space
of Lie group at identity, and there is no unique way of choosing of
the Lie algebra basis. In this work, the three linearly independent
basis of $\mathfrak{su}(2)$ will be chosen to satisfy
\begin{equation}
\mathfrak{g}_{i}\mathfrak{g}_{j}=-\delta_{ij}+\epsilon_{ijk}\mathfrak{g}_{k}\label{eq:Lie_algebra}
\end{equation}
By comparing Eq.\ref{eq:quaternion} and Eq.\ref{eq:Lie_algebra},
there is a clear one-to-one map between the three imaginary number
of quaternion (i.e. $\hat{i},\hat{j,}\hat{k}$) and the three basis
of $\mathfrak{su}(2)$ (i.e. $\mathfrak{g}_{1},\mathfrak{g}_{2},\mathfrak{g}_{3}$),
thus they adopt the same matrix representations as shown in Table
\ref{table:math}. This also implies that any pure unit quaternion
$\hat{n}$ can be mapped to a Lie algebra $\mathfrak{n}\in\mathfrak{su}(2)$
as
\begin{equation}
\hat{n}=n_{1}\hat{i}+n_{2}\hat{j}+n_{3}\hat{k}\leftrightarrow\mathfrak{n}=n_{1}\mathfrak{\mathfrak{g}}_{1}+n_{2}\mathfrak{\mathfrak{g}}_{2}+n_{3}\mathfrak{\mathfrak{g}}_{3}\label{eq:n_to_lie}
\end{equation}
 In Section \ref{sec:Topological}, the Berry connection and Berry
curvature are evaluated in quaternion, and the results can be mapped
to $\mathfrak{su}(2)$ using Eq.\ref{eq:n_to_lie}.

As shown in Eq.\ref{eq:exp_q}, a unit quaternion can be generated
from a pure unit quaternion $\hat{n}$ through exponential function.
On the other hand, a Lie group element $g$ can be generated from
a Lie algebra $\mathfrak{g}$ through exponential map (i.e. the one-parameter
group). As a result, there is also a one-to-one map between a unit
quaternion $e^{\hat{n}\phi}\in\mathbb{H}^{1}$ and a group element
$\text{Exp}(\mathfrak{n}\phi)\in SU(2)$. This is the well known isomorphism
of $\mathbb{H}^{1}\simeq S^{3}\simeq SU(2)$.

\section{Conical intersection models\label{sec:Conical-intersection}}

The derivations of conical intersection model Hamiltonians based on
general perturbation theory using crude adiabatic states have been
discussed in details in the book chapter by Yarkony.\cite{domckeConicalIntersectionsElectronic2004}
Starting from these model Hamiltonians, we will first show how they
can be written in terms of geometric algebra in Section \ref{subsec:modelH}.
We will then discuss how the corresponding eigenfunctions can be easily
obtained by using geometric algebra to explore the rotational symmetry
of the Hamiltonian in Section \ref{subsec:Solve-modelH}.

\subsection{Model Hamiltonians\label{subsec:modelH}}

\subsubsection{$T^{2}=0$ conical intersection}

With no time reversal symmetry, the Hamiltonian of a two-level system
is a complex Hermitian matrix. Therefore, the model Hamiltonian in
the vicinity of $T^{2}=0$ conical intersection takes the form of
\begin{equation}
H^{3d}=\left(\begin{array}{cc}
gz & h^{r}x-ih^{i}y\\
h^{r}x+ih^{i}y & -gz
\end{array}\right)=gz\cdot\sigma_{3}+h^{r}x\cdot\sigma_{1}+h^{i}y\cdot\sigma_{2}
\end{equation}
with $x,y,z$ define the three orthogonal directions in the branching
plane that lift the degeneracy linearly. We will use the following
scaled Cartesian coordinate: Let $R=((gz)^{2}+(h^{r}x)^{2}+(h^{i}y)^{2})^{1/2}$,
and define $r_{x}=\frac{1}{R}h^{r}x$, $r_{y}=\frac{1}{R}h^{i}y$,
$r_{z}=\frac{1}{R}gz$ such that $\vec{r}=(r_{x},r_{y},r_{z})$ is
a unit vector on $S^{2}$. Furthermore, by using the one-to-one mapping
between $\sigma_{i}$ and\textbf{ $\vec{e}_{i}\in Cl(3)$} in Table
\ref{table:math}, the above Hamiltonian can be written in terms of
geometric algebra as
\begin{equation}
H^{3d}=R\cdot\sum_{i\in\{x,y,z\}}r_{i}\vec{e}_{i}\label{eq:H_3D}
\end{equation}
Since Pauli matrices $\text{\ensuremath{\vec{e}_{i}\simeq\sigma_{i}}}$
all have the same eigenvalues $\{+1,-1\}$, the amount that the degeneracy
is lifted only depends on the radius $R$. Therefore, the Hamiltonian
in Eq.\ref{eq:H_3D} is of SO(3) symmetry.

\subsubsection{$T^{2}=-1$ conical intersection}

A detailed explanation about the model Hamiltonian of $T^{2}=-1$
conical intersection can be found in a series of work by Matsika and
Yarkony\cite{domckeConicalIntersectionsElectronic2004,matsikaSpinOrbitCoupling2002,matsikaEffectsSpinorbitCoupling2001}.
The time reversal operator in these works takes the form as $\hat{T}=(i\sigma_{2}\mathcal{K})\otimes I$
(note $\mathcal{K}$ is the complex conjugation operator). In order
to relate to geometric algebra in later discussions, we permuted their
ordering of rows and columns as follows, which is equivalent to writing
the time reversal operator as $\hat{T}=I\otimes(i\sigma_{2}\mathcal{K})$
\begin{equation}
H^{5d}=\left(\begin{array}{cccc}
gz & 0 & h^{r}x+ih^{i}y & t^{r}u+it^{i}v\\
0 & gz & -t^{r}u+it^{i}v & h^{r}x-ih^{i}y\\
h^{r}x-ih^{i}y & -t^{r}u-it^{i}v & -gz & 0\\
t^{r}u-it^{i}v & h^{r}x+ih^{i}y & 0 & -gz
\end{array}\right)\label{eq:H5d_detail}
\end{equation}
Similar to the $T^{2}=0$ case, we use scaled coordinates by letting
$R=((gz)^{2}+(h^{r}x)^{2}+(h^{i}y)^{2}+(t^{r}u)^{2}+(t^{i}v)^{2})^{1/2}$,
and define $r_{z}=\frac{1}{R}gz$, $r_{x}=\frac{1}{R}h^{r}x$, $r_{y}=\frac{1}{R}h^{i}y$,
$r_{u}=\frac{1}{R}t^{r}u$, $r_{v}=\frac{1}{R}t^{i}v$. The coordinates
$\vec{r}=(r_{z},r_{x},r_{y},r_{u},r_{v})$ thus define a point on
$S^{4}$. This transforms the Hamiltonian into the form of
\begin{equation}
H^{5d}=R\sum_{i\in\{xyzuv\}}r_{i}\Gamma_{i}
\end{equation}
where
\begin{equation}
\begin{array}{c}
\Gamma_{z}=\sigma_{z}\otimes I,\text{\text{ }}\Gamma_{x}=\sigma_{x}\otimes I\\
\Gamma_{y}=\left(-i\sigma_{y}\right)\otimes\left(-i\sigma_{z}\right),\text{\text{ }}\Gamma_{u}=\left(-i\sigma_{y}\right)\otimes\left(-i\sigma_{y}\right), \\\text{\text{ }}\Gamma_{v}=\left(-i\sigma_{y}\right)\otimes\left(-i\sigma_{x}\right)
\end{array}\label{eq:Gamma_matrix}
\end{equation}
One can verify that the five $\Gamma$-matrices satisfy $\Gamma_{i}\Gamma_{j}+\Gamma_{j}\Gamma_{i}=0$,
which is the defining properties of geometric algebra generators in
Eq.\ref{eq:clifford}. Therefore, the five $\Gamma$-matrices form
the matrix representations of $Cl(5)$. This again allows us to rewrite
the Hamiltonian in terms of geometric algebra as
\begin{equation}
H^{5d}=R\sum_{i\in\{xyzuv\}}r_{i}\vec{e}_{i}\label{eq:H_5D}
\end{equation}
Because the five $\Gamma$-matrices all have the same eigenvalues
$\{+1,+1,-1-1\}$, the lift of degeneracy only depends on radius $R,$
making the Hamiltonian of SO(5) symmetry.

The above equations are all written in complex numbers. Using matrix
representations of quaternions in Table \ref{table:math}, the $4\times4$
complex model Hamiltonian matrix in Eq.\ref{eq:H5d_detail} can be
further simplified as $2\times2$ quaternion matrix as
\begin{equation}
H^{5d}=\left(\begin{array}{cc}
gz\cdot\hat{1} & h^{r}\hat{1}-h^{i}\hat{k}-t^{r}\hat{j}-t^{i}\hat{i}\\
h^{r}\hat{1}+h^{i}\hat{k}+t^{r}\hat{j}+t^{i}\hat{i} & -gz\cdot\hat{1}
\end{array}\right)
\end{equation}
where $\hat{1}$ is the identity quaternion. Similarly, the $4\times4$
complex matrix representations of $\vec{e}_{i}$ of $Cl(5)$ in Eq.\ref{eq:Gamma_matrix}
can also be simplified as $2\times2$ quaternion matrices
\begin{equation}
\begin{array}{c}
\vec{e}_{z}=\left(\begin{array}{cc}
\hat{1}\\
 & -\hat{1}
\end{array}\right),\text{\text{ }}\vec{e}_{x}=\left(\begin{array}{cc}
 & \hat{1}\\
\hat{1}
\end{array}\right)\\
\vec{e}_{y}=\left(\begin{array}{cc}
 & -\hat{k}\\
\hat{k}
\end{array}\right),\text{\text{ }}\vec{e}_{u}=\left(\begin{array}{cc}
 & -\hat{j}\\
\hat{j}
\end{array}\right),\text{\text{ }}\vec{e}_{v}=\left(\begin{array}{cc}
 & -\hat{i}\\
\hat{i}
\end{array}\right)
\end{array}\label{eq:Cl(5)}
\end{equation}
This shows that the model Hamiltonian of $T^{2}=-1$ conical intersection
is in fact a Hermitian quaternion matrix, which is consistent with
what is known from other $T^{2}=-1$ system.\textcolor{red}{{} }\cite{avronTopologicalInvariantsFermi1988}\textcolor{red}{{} }

\subsection{Solve the eigenfunctions through geometric algebra\label{subsec:Solve-modelH}}

\begin{figure*}[t]
	\begin{center}
		\includegraphics[width=4.5in]{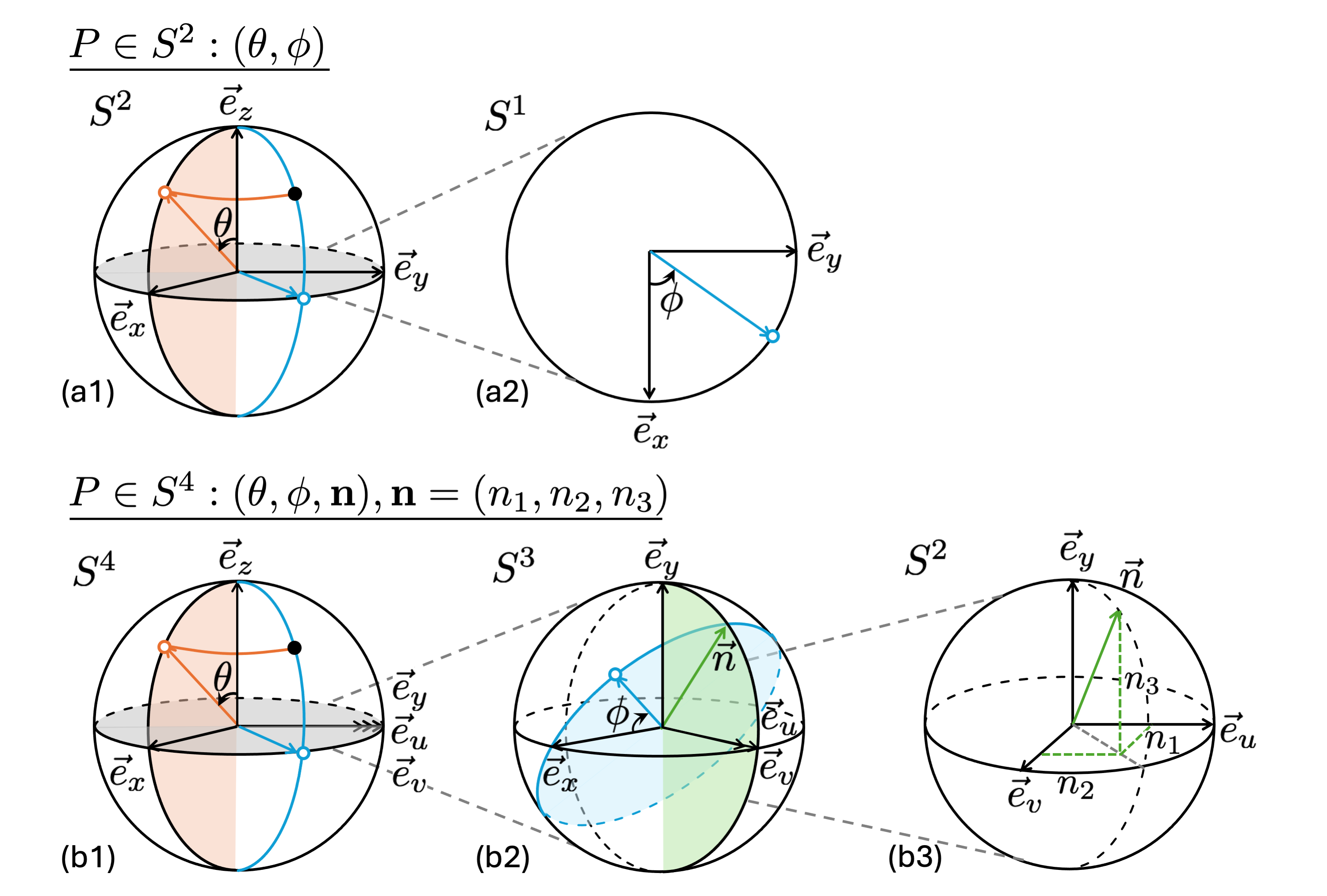}
		\caption{\label{fig:coors}
Illustrations of coordinate
systems used in this work. As discussed in the main text, the wavefunctions
only depend on the angular coordinates, so the radial coordinate $R$ is
not shown. $S^{n}$ on the upper left for each panel indicates the
dimension of the sphere that is shown. (a1)-(a2) illustrate coordinates
$(\theta,\phi)$ for studying $T^{2}=0$ conical intersection, which
are consistent with the standard spherical coordinate. (b1)-(b3) illustrate
coordinates $(\theta,\phi,\mathbf{n})$ for studying $T^{2}=-1$ conical
intersection.
		}
		\end{center}
\end{figure*}

As mentioned in Section \ref{subsec:modelH}, in the scaled Cartesian
coordinates, $H^{3d}$ in Eq.\ref{eq:H_3D} has SO(3) symmetry while
$H^{5d}$ in Eq.\ref{eq:H_5D} has SO(5) symmetry, and their eigenvalues
only depend on the radius $R$. Geometric algebra provides a way to
solve the corresponding eigenvectors by making use of this rotational
symmetry.

In general, the model Hamiltonian is parameterized by the position
in the branching space $\vec{R}$, and the Schrodinger equation takes
the form of
\begin{equation}
H[\vec{R}]\Psi[\vec{R}]=E(\vec{R})\Psi[\vec{R}]
\end{equation}
where we use square bracket to denote parametrization. The symmetry
of the Hamiltonian is characterized by its symmetry group $G$. Let
$g$ be a group element of the symmetry group $G$ such that $\vec{R}^{\prime}=g(\vec{R})$.
The Schrodinger equation at $\vec{R}^{\prime}$ can be written as
\begin{equation}
H[g(\vec{R})]\Psi[g(\vec{R})]=E(g(\vec{R}))\Psi[g(\vec{R})]
\end{equation}
On the other hand, $g$ also induces a coordinate transformation,
which transforms the Schrodinger equation as
\begin{equation}
g(H[\vec{R}])\cdot g(\Psi[\vec{R}])=E(\vec{R})\cdot g(\Psi[\vec{R}])
\end{equation}
The fact that $G$ is a symmetry group of the Hamiltonian implies
that for any $g\in G$
\begin{equation}
E(g(\vec{R}))=E(\vec{R}),\text{\text{ }}g(\Psi[\vec{R}])\simeq\Psi[g(\vec{R})]\label{eq:symmetry}
\end{equation}
where $\simeq$ indicates that the two wavefunctions are equivalent
up to a gauge transformation (called ``gauge-equivalent'' for short).
In the case that $G$ is a rotation group, the explicit forms of $g(H[\vec{R}])$
and $g(\Psi[\vec{R}])$ depend on how they transform under rotation
respectively. For $g(H[\vec{R}])$, Eq.\ref{eq:H_3D} and Eq.\ref{eq:H_5D}
show that the model Hamiltonian transforms as \emph{vectors }from
the perspective of geometric algebra, i.e. Eq.\ref{eq:vector_rotation}.
In terms of $g(\Psi[\vec{R}])$, because $\left\{ R(\theta)H[\vec{R}]R(-\theta)\right\} \cdot\left\{ R(\theta)\Psi[\vec{x}]\right\} =E(\vec{x})\left\{ R(\theta)\Psi[\vec{x}]\right\} $,
the wavefunction transforms as\emph{ spinor} instead, i.e. Eq.\ref{eq:spinor_rotation}.

Eq.\ref{eq:symmetry} shows that if we know the eigenfunction of $\hat{H}$
at one location $\Psi[\vec{R}]$, we can then use the group element
$g\in G$ to ``transplant'' the eigenfunction to a different location
$\vec{R}^{\prime}=g(\vec{R})$. Below we will first demonstrate this
idea and highlight the role of geometric algebra using the simpler
case of $T^{2}=0$. We will then show how the idea can be generalized
to $T^{2}=-1$.

The fact that the radius $R$ appears as a prefactor in Eq.\ref{eq:H_3D}
and Eq.\ref{eq:H_5D} suggests that the corresponding eigenfunctions
are independent of $R$. Therefore, the discussions in Section \ref{subsec:solve_3D}
and \ref{subsec:solve_5D} will restrict to coordinates on $S^{2}$
for $T^{2}=0$, and $S^{4}$ for $T^{2}=-1$.

\subsubsection{$T^{2}=0$ conical intersection\label{subsec:solve_3D}}

The spherical coordinate system we used for studying $T^{2}=0$ conical
intersection is illustrated in Figure \ref{fig:coors}(a1)-(a2). The
north pole $\mathcal{N}$ of $S^{2}$ will be defined as the point
where it intersects with the z-axis. The standard polar coordinate
$(\theta,\phi)$ will be used to specify locations on $S^{2}$, where
$\theta$ is the latitude and $\phi$ is the longitude. They are related
to the scaled Cartesian coordinates as $r_{z}=\cos\theta$, $r_{x}=\sin\theta\cos\phi$
and $r_{y}=\sin\theta\sin\phi$. In order to use $\Psi[g(\vec{r}_{0})]\simeq g(\Psi[\vec{r}_{0}])$
from Eq.\ref{eq:symmetry}, we will choose the reference location
$\vec{r}_{0}$ to be the north pole, and choose the corresponding
eigenfunctions $\Psi[\mathcal{N}]$ as

\begin{equation}
\Psi^{+}[\mathcal{N}]=\left(\begin{array}{c}
1\\
0
\end{array}\right),\text{\text{ }}\Psi^{-}[\mathcal{N}]=\left(\begin{array}{c}
0\\
1
\end{array}\right)
\end{equation}
where $\Psi^{+}$ is the upper state and $\Psi^{-}$ is the lower
state. In order to find the $\Psi(\theta,\phi)$, we need the group
element $g$ that rotates the north pole to $(\theta,\phi)$. This
can be decomposed as the following two successive rotations. First,
rotate within the the zx-plane (defined by the bivector $\check{\beta}_{zx}$)
by $\theta$, see Figure \ref{fig:coors}(a1). Using the matrix representation
of bivectors $\check{\beta}_{zx}$ in Table \ref{table:math}, the
corresponding rotor (Eq.) is then defined as
\begin{equation}
R_{zx}(\theta)=\cos\frac{\theta}{2}-\sin\frac{\theta}{2}\check{\beta}_{zx}=\left(\begin{array}{cc}
\cos\frac{\theta}{2} & -\sin\frac{\theta}{2}\\
\sin\frac{\theta}{2} & \cos\frac{\theta}{2}
\end{array}\right)
\end{equation}
One then needs to rotate within the plane of $\check{\beta}_{xy}$
by $\phi$ (Figure \ref{fig:coors}(a2)), and the rotor is defined
as
\begin{equation}
R_{xy}(\phi)=\cos\frac{\phi}{2}-\sin\frac{\phi}{2}\check{\beta}_{xy}=\left(\begin{array}{cc}
e^{-i\frac{\phi}{2}}\\
 & e^{i\frac{\phi}{2}}
\end{array}\right)
\end{equation}
The rotor for the desired rotation is thus $g=R_{xy}(\phi)R_{zx}(\theta)$.
Because eigenfunction transforms as spinor (Eq.\ref{eq:spinor_rotation}),
$\Psi(\theta,\phi)$ is gauge equivalent to $g(\Psi[\mathcal{N}])$
as
\begin{equation}
\Psi[\theta,\phi]=R_{xy}(\phi)R_{zx}(\theta)\Psi[\mathcal{N}]\label{eq:rotor_3D}
\end{equation}
This leads to the following results for upper and lower state respectively
\begin{equation}
\Psi^{+}[\theta,\phi]=\left(\begin{array}{c}
\cos\frac{\theta}{2}e^{-i\frac{\phi}{2}}\\
\sin\frac{\theta}{2}e^{i\frac{\phi}{2}}
\end{array}\right),\text{\text{ }}\Psi^{-}[\theta,\phi]=\left(\begin{array}{c}
-\sin\frac{\theta}{2}e^{-i\frac{\phi}{2}}\\
\cos\frac{\theta}{2}e^{i\frac{\phi}{2}}
\end{array}\right)\label{eq:solution_3D}
\end{equation}

The above idea of solving the eigenfunctions using SO(3) symmetry
is often implicitly used for studying Bloch sphere ($T^{2}=0$) in
quantum computing,\cite{nielsenQuantumComputationQuantum2000} which
are often presented through Pauli matrices without direct mentioning
$Cl(3)$ geometric algebra. The main purpose of this section is to
highlight the different roles of vectors and bivectors in the derivations,
which will become important in the next section where vectors and
bivectors of $Cl(5)$ no longer have duality.

\subsubsection{$T^{2}=-1$ conical intersection\label{subsec:solve_5D}}

Matsika et.al\cite{matsikaSpinOrbitCoupling2002} showed how to directly
solve the eigenfunctions of Hamiltonian Eq.\ref{eq:H_5D} in complex
numbers by considering the $T^{2}=-1$ conical intersections as two
coupled $T^{2}=0$ ones.\cite{domckeConicalIntersectionsElectronic2004}
The procedure is more complicated than $T^{2}=0$ due to the fact
that the model Hamiltonian is a $4\times4$ matrix. In this section,
we will show that the combination of geometric algebra and quaternion
number provides a simple way to solve the eigenfunctions that closely
resembles the procedure in Section \ref{subsec:solve_3D}.

The coordinate we use to study $T^{2}=-1$ conical intersection is
illustrated in Figure \ref{fig:coors}(b1)-(b3). The north pole $\mathcal{N}$
of $S^{4}$ is again defined as the point where it intersects with
$\vec{e}_{z}$. To specify a point on $S^{4}$, the latitude $\theta$
is defined in the zx-plane resembling the $S^{2}$ situation (Figure
\ref{fig:coors}(b1)), but the longitude degree of freedom needs to
be defined along $S^{3}$-equator instead of $S^{1}$. Within $S^{3}$-equator,
any point can be rotated from $\vec{e}_{x}$ by a certain angle $\phi$
(Figure \ref{fig:coors}(b2)). The corresponding rotation plane is
spanned by $\vec{e}_{x}$ and a unit vector that is perpendicular
to $\vec{e}_{x}$. Such unit vector can be expressed as $\vec{n}=n_{3}\vec{e}_{y}+n_{2}\vec{e}_{u}+n_{1}\vec{e}_{v}$,
$\sum_{i=1}^{3}n_{i}^{2}=1$ (Figure \ref{fig:coors}(b3)). Overall,
this shows that any point on $S^{4}$ can be specified using coordinates
$(\theta,\phi,\mathbf{n}=\{n_{1},n_{2},n_{3}\})$, which is related
to the scaled Cartesian coordinates as: $r_{z}=\cos\theta$, $r_{x}=\sin\theta\cos\phi$,
$r_{y}=\sin\theta\sin\phi n_{3}$, $r_{u}=\sin\theta\sin\phi n_{2}$,$r_{v}=\sin\theta\sin\phi n_{1}$.

Similar to how we solve $T^{2}=0$, we will start with the eigenfunctions
at the north pole in quaternion numbers, i.e.

\begin{equation}
\Psi^{+}[\mathcal{N}]=\left(\begin{array}{c}
\hat{1}\\
\hat{0}
\end{array}\right),\text{\text{ }}\Psi^{-}[\mathcal{N}]=\left(\begin{array}{c}
\hat{0}\\
\hat{1}
\end{array}\right)
\end{equation}
To construct rotor that rotates the north pole to $(\theta,\phi,\mathbf{n})$,
we need to use geometric algebra $Cl(5)$ in Eq.\ref{eq:Cl(5)}. First,
similar to $T^{2}=0$, rotate within the plane of $\check{\beta}_{zx}$
plane by $\theta$ (Figure \ref{fig:coors}(b1)), except that $\check{\beta}_{zx}$
is a bivector in 5D. To find matrix representation of $\check{\beta}_{zx}$
in 5D, substitute $Cl(5)$ generators in Eq.\ref{eq:Cl(5)} into the
definition of bivector in Eq.\ref{eq:bivector}
\begin{equation}
\check{\beta}_{zx}=\vec{e}_{z}\vec{e}_{x}=\left(\begin{array}{cc}
\hat{1}\\
 & -\hat{1}
\end{array}\right)\left(\begin{array}{cc}
 & \hat{1}\\
\hat{1}
\end{array}\right)=\left(\begin{array}{cc}
 & \hat{1}\\
-\hat{1}
\end{array}\right)
\end{equation}
The corresponding rotor can thus be expressed as
\begin{equation}
R_{zx}(\theta)=\cos\frac{\theta}{2}-\sin\frac{\theta}{2}\check{\beta}_{zx}=\left(\begin{array}{cc}
\cos\frac{\theta}{2}\hat{1} & -\sin\frac{\theta}{2}\hat{1}\\
\sin\frac{\theta}{2}\hat{1} & \cos\frac{\theta}{2}\hat{1}
\end{array}\right)
\end{equation}
Second, rotate within the plane of $\check{\beta}_{xn}$ by $\phi$
(Figure \ref{fig:coors}(b2)). The vector $\vec{n}$ can be represented
in terms of geometric algebra as (using Eq.\ref{eq:Cl(5)})
\begin{equation}
\vec{n}=n_{3}\vec{e}_{y}+n_{2}\vec{e}_{u}+n_{1}\vec{e}_{v}=\left(\begin{array}{cc}
 & -\left(n_{3}\hat{k}+n_{2}\hat{j}+n_{1}\hat{i}\right)\\
n_{3}\hat{k}+n_{2}\hat{j}+n_{1}\hat{i}
\end{array}\right)=\left(\begin{array}{cc}
 & -\hat{n}\\
\hat{n}
\end{array}\right)\label{eq:vec_n}
\end{equation}
and the bivector $\check{\beta}_{xn}$ can again be computed from
Eq.\ref{eq:bivector} as
\begin{equation}
\check{\beta}_{xn}=\vec{e}_{x}\vec{n}=\left(\begin{array}{cc}
 & \hat{1}\\
\hat{1}
\end{array}\right)\left(\begin{array}{cc}
 & -\hat{n}\\
\hat{n}
\end{array}\right)=\left(\begin{array}{cc}
\hat{n}\\
 & -\hat{n}
\end{array}\right)\label{eq:bivector_xn}
\end{equation}
Since $\hat{n}$ is a unit pure quaternion, using its property Eq.\ref{eq:exp_q},
the corresponding rotor can be defined as

\begin{equation}
R_{xn}(\phi)=\cos\frac{\phi}{2}-\sin\frac{\phi}{2}\check{\beta}_{xn}=\left(\begin{array}{cc}
e^{-\hat{n}\frac{\phi}{2}}\\
 & e^{\hat{n}\frac{\phi}{2}}
\end{array}\right)
\end{equation}
By combining these two steps together, the desired rotor is $R_{xn}(\phi)R_{zx}(\theta)$.
Similar to Eq.\ref{eq:rotor_3D}, the eigenfunctions at $(\theta,\phi,\mathbf{n})$
can now be computed as
\begin{equation}
\Psi[\theta,\phi,\mathbf{n}]=R_{xn}(\phi)R_{zx}(\theta)\Psi[\mathcal{N}]
\end{equation}
which leads to the following eigenfunctions
\begin{equation}
\Psi^{+}[\theta,\phi,\mathbf{n}]=\left(\begin{array}{c}
\cos\frac{\theta}{2}e^{-\hat{n}\frac{\phi}{2}}\\
\sin\frac{\theta}{2}e^{\hat{n}\frac{\phi}{2}}
\end{array}\right),\text{\text{ }}\Psi^{-}[\theta,\phi,\mathbf{n}]=\left(\begin{array}{c}
-\sin\frac{\theta}{2}e^{-\hat{n}\frac{\phi}{2}}\\
\cos\frac{\theta}{2}e^{\hat{n}\frac{\phi}{2}}
\end{array}\right)\label{eq:solution_5D}
\end{equation}

Eq.\ref{eq:solution_5D} provides the eigenfunction for each state
as $2\times1$ quaternion vector. This can be translated back to $4\times2$
complex vector using Eq.\ref{eq:exp_q_matrix}. Note that since each
quaternion number corresponds to $2\times2$ complex matrix, the two-fold
Kramer degeneracy is automatically accounted for.

\section{Topological properties of $T^{2}=-1$ conical intersection\label{sec:Topological}}

Based on the eigenfunctions in Eq.\ref{eq:solution_5D}, we will study
the geometric and topological properties of the $T^{2}=-1$ conical
intersection by drawing connection with the properties of Yang monopole.
The conical intersection is a point singularity when viewed from the
corresponding branching space. Based on gauge theory,\cite{nakaharaGeometryTopologyPhysics2003}
the coordinate atlas should be composed of at least two charts, and
a common choice contains a northern chart that excludes $-z$ axis
plus a southern chart that excludes $+z$ axis. The non-trivial topology
of conical intersection prevents the adoption of a global gauge, and
local gauge needs to be chosen for each chart. The two local gauges
on the northern and southern charts are then ``glued'' together
according to the ``transition function'' at the overlapping region,
e.g. the equator.

The gauge theory of Dirac monopole has two features:\cite{wuConceptNonintegrablePhase1975}(1)
Because the gauge group U(1) is isomorphic to the equator $S^{1}$,
local gauges can be chosen such that the transition function $T$
at $\phi\in S^{1}$ naturally maps to the gauge group as $e^{i\phi}$
or $e^{-i\phi}$. (2) The phase factor $\Phi$ from point $P:(R,\theta,\phi)$
to $P+dP:(R+dR,\theta+d\theta,\phi+d\phi)$ relates to the transition
function $t$ as $\Phi=\left(t_{P+dP}t_{P}^{-1}\right)^{p(\theta)}$
with $p(\theta)=\frac{1}{2}(1-\cos\theta)$ on the northern chart,
and as $\Phi=\left(t_{P+dP}^{-1}t_{P}\right)^{1-p(\theta)}$on the
southern chart. Yang constructs his SU(2) monopole by imposing two
similar condition:\cite{yangGeneralizationDiracMonopole1978} (1)
the transition function along the $S^{3}$-equator is defined through
its isomorphism to the gauge group $S^{3}\simeq SU(2)$ and (2) the
phase factor from $P$ to $P+dP$ on the two charts are related to
the transition function in the same way as the Dirac monopole. Because
the phase factor is line integration of gauge potential, property
(2) can also be written in its differential form that relates transition
function with gauge potential.

Below we will show that $T^{2}=-1$ conical intersection indeed satisfies
the above two defining conditions thus is a Yang monopole. Based on
the relations between Dirac and Yang monopoles, we compare the properties
of $T^{2}=0$ and $T^{2}=-1$ conical intersections in Table \ref{table:3Dvs5D}.
The transition function and Berry connection entries are explained
in Section \ref{subsec:Yang-monopole} and \ref{subsec:constant_BD}.
The Berry curvature and Chern number entries are explained in Section
\ref{subsec:Berry-curvature}.

\begin{table*}[t]
\caption{Comparison of the properties between molecular conical intersections in magnetic field ($T^{2}=0$) versus those not in magnetic field and have odd number of electrons ($T^{2}=-1$). In the last four rows, $\pm$ on the left hand side of the equations indicate properties of upper and lower energy levels respectively.}
\label{table:3Dvs5D}
{
\begin{tabular}{|c|c|c|}
\hline
Molecular & In magnetic field & No magnetic field\tabularnewline
conical intersection &  & odd electron number\tabularnewline
\hline
\hline
Time reversal symmetry & $T^{2}=0$ & $T^{2}=-1$\tabularnewline
\hline
Model $H$ in scaled coordinates & $H=\sum_{i=1}^{3}r_{i}\vec{e}_{i}$ & $H=\sum_{i=1}^{5}r_{i}\vec{e}_{i}$\tabularnewline
\hline
Monopole type & Dirac monopole & Yang monopole\tabularnewline
\hline
Number system & \textcolor{black}{Complex number} & \textcolor{black}{Quaternion number}\tabularnewline
\hline
Branching space & 3D & 5D\tabularnewline
\hline
Rotational symmetry  & SO(3) & SO(5)\tabularnewline
\hline
Equator $\simeq$ Gauge group & $S^{1}\simeq\mathbb{C}^{1}\simeq U(1)$ & $S^{3}\simeq\mathbb{H}^{1}\simeq SU(2)$\tabularnewline
\hline
Transition function & $t_{\pm}^{SN}=e^{\pm i\phi}$ & $t_{\pm}^{SN}=e^{\pm\hat{n}\phi}$\tabularnewline
\hline
Berry connection & $A_{R}=0,A_{\theta}=0$ & $A_{R}=0,A_{\theta}=0$\tabularnewline
 & $A_{\phi}^{\pm,N}=\pm i\frac{1-\cos\theta}{2}$ & $A_{\phi^{\prime}}^{\pm,N}(S_{pq}^{1})=\hat{b}_{\pm}(S_{pq}^{1})\frac{1-\cos\theta}{2}$\tabularnewline
 & $A_{\phi}^{\pm,S}=\mp i\frac{1+\cos\theta}{2}$ & $A_{\phi^{\prime}}^{\pm,S}(S_{pq}^{1})=\hat{b}_{\mp}(S_{pq}^{1})\frac{1+\cos\theta}{2}$\tabularnewline
\hline
Berry curvature & $F_{12}^{\pm}=\pm\frac{1}{2}i$ & Eq.\ref{eq:F_standard}\tabularnewline
\hline
Chern number & $C_{1}^{\pm}=\pm\frac{1}{2}$ & $C_{2}^{\pm}=\pm\frac{1}{2^{2}}$\tabularnewline
 & $\frac{1}{2}\oint\sum_{\mu\nu}F_{\mu\nu}dx^{\mu\nu}=4\pi iC_{1}$ & $\frac{1}{4!}\oint{\displaystyle \sum_{\mu\nu\alpha\beta n}}F_{\mu\nu}^{n}F_{\alpha\beta}^{n}dx^{\mu\nu\alpha\beta}=\frac{8\pi^{2}}{3}C_{2}$\tabularnewline
\hline
\end{tabular}

}
\end{table*}

\subsection{$T^{2}=-1$ conical intersection is a Yang monopole\label{subsec:Yang-monopole}}

As mentioned above, the transition functions for both $T^{2}=0$ and
$T^{2}=-1$ are defined through the isomorphism of the equator geometry
with the gauge group, which associates a group element $g_{l}$ with
every point in the equator as
\begin{equation}
g_{l}=\left\{ \begin{array}{cc}
l=\phi\in S^{1}\leftrightarrow e^{i\phi}\in U(1), & \text{ for }T^{2}=0\\
l=\{\phi,\mathbf{n}\}\in S^{3}\leftrightarrow e^{\hat{n}\phi}\in SU(2) & \text{ for }T^{2}=-1
\end{array}\right.\label{eq:g_transition}
\end{equation}
The solutions in Eq.\ref{eq:solution_3D} for $T^{2}=0$ and Eq.\ref{eq:solution_5D}
for $T^{2}=-1$ can be combined into the following form
\begin{equation}
\Psi^{+}[\theta,l]=\left(\begin{array}{c}
\cos\frac{\theta}{2}g_{l}^{-1/2}\\
\sin\frac{\theta}{2}g_{l}^{1/2}
\end{array}\right),\text{ }\Psi^{-}[\theta,l]=\left(\begin{array}{c}
-\sin\frac{\theta}{2}g_{l}^{-1/2}\\
\cos\frac{\theta}{2}g_{l}^{1/2}
\end{array}\right)
\end{equation}
Because Berry connection is gauge-dependent, only a specific choice
of local gauges can produce the desired formula that matches with
the gauge potential in Yang's paper. The required local gauge can
be chosen as follows. On the northern chart, because the longitude
parameter $l$ is not well defined at the north pole ($\theta=0$),
we adopt the following local gauges
\begin{equation}
\Psi^{+,N}[\theta,l]=\left(\begin{array}{c}
\cos\frac{\theta}{2}\\
\sin\frac{\theta}{2}g_{l}
\end{array}\right),\text{ }\Psi^{-,N}[\theta,l]=\left(\begin{array}{c}
-\sin\frac{\theta}{2}g_{l}^{-1}\\
\cos\frac{\theta}{2}
\end{array}\right)\label{eq:north_chart}
\end{equation}
On the southern chart, $l$ is not well-defined at the south pole
($\theta=\pi$), and we adopt the following gauge for $\Psi^{+}$
and $\Psi^{-}$ respectively
\begin{equation}
\Psi^{+,S}[\theta,l]=\left(\begin{array}{c}
\cos\frac{\theta}{2}g_{l}^{-1}\\
\sin\frac{\theta}{2}
\end{array}\right),\text{ }\Psi^{-,S}[\theta,l]=\left(\begin{array}{c}
-\sin\frac{\theta}{2}\\
\cos\frac{\theta}{2}g_{l}
\end{array}\right)\label{eq:south_chart}
\end{equation}
By comparing Eq.\ref{eq:north_chart} and Eq.\ref{eq:south_chart},
the transition function (defined to act on the right $\Psi^{N}=\Psi^{S}t^{SN}$
) is related to $g_{l}$ as
\begin{equation}
t_{+}^{SN}(l)=g_{l},\text{ }t_{-}^{SN}(l)=g_{l}^{-1}
\end{equation}
This is consistent with the first defining condition of Yang monopole,
i.e. the transition function reflects the isomorphism between gauge
group and equator geometry.

The Berry connection along a particular direction $\frac{\partial}{\partial x_{\mu}}$
is defined as\cite{nakaharaGeometryTopologyPhysics2003}
\begin{equation}
A_{\mu}=\langle\Psi\rvert\frac{\partial}{\partial x_{\mu}}\Psi\rangle\label{eq:Berry_connection}
\end{equation}
The eigenfunctions in Eq.\ref{eq:north_chart} and Eq.\ref{eq:south_chart}
have no dependency on the radial coordinate $R,$ thus the Berry connection
is zero along the radial direction, i.e.
\begin{equation}
A_{R}^{+/-,N/S}=0\label{eq:A_R}
\end{equation}
Using the fact that $g_{l}$ defined in Eq.\ref{eq:g_transition}
does not depend on $\theta$, some simple derivations will show that
the Berry connection along $\theta$ is zero as well
\begin{equation}
A_{\theta}^{+/-,N/S}=0\label{eq:A_theta}
\end{equation}
Lastly, for any direction $\frac{\partial}{\partial\lambda}$ within
the longitude degree of freedom, the Berry connection is related to
$\frac{\partial g_{l}}{\partial\lambda}.$ By defining
\begin{equation}
D_{\lambda}\left(l\right)=-\frac{\partial g_{l}^{-1}}{\partial\lambda}g_{l},\text{ }B_{\lambda}(l)=-\frac{\partial g_{l}}{\partial\lambda}g_{l}^{-1}\label{eq:A_factor}
\end{equation}
one can show that the Berry connection along any longitude direction
$\frac{\partial}{\partial\lambda}$ takes the form of
\begin{equation}
\begin{array}{c}
A_{\lambda}^{+,N}(\theta,l)=\frac{1-\cos\theta}{2}D_{\lambda}(l),\text{ }A_{\lambda}^{+,S}(\theta,l)=\frac{1+\cos\theta}{2}B_{\lambda}(l)\\
A_{\lambda}^{-,N}(\theta,l)=\frac{1-\cos\theta}{2}B_{\lambda}(l),\text{ }A_{\lambda}^{-,S}(\theta,l)=\frac{1+\cos\theta}{2}D_{\lambda}(l)
\end{array}\label{eq:A_longitude}
\end{equation}
Eqs. \ref{eq:A_R}, \ref{eq:A_theta} and \ref{eq:A_longitude} completely
specify the Berry connection in the branching space. Note that the
derivations from Eq.\ref{eq:g_transition} through Eq.\ref{eq:A_longitude}
are applicable to both $T^{2}=0$ and $T^{2}=-1$ conical intersections.
In particular, by substitute $g_{\phi}=e^{i\phi}$ into Eq.\ref{eq:A_factor}
for $T^{2}=0$, one can show that $D_{\phi}=i,B_{\phi}=-i$, leading
to the Berry connection formula listed in Table \ref{table:3Dvs5D}.

Yang has mathematically proved that there are two and only two types
of SU(2) monopole that has SO(5) symmetry, denoted as $\alpha$ and
$\beta$ type respectively. By comparing Eq. \ref{eq:A_R}, \ref{eq:A_theta}
and \ref{eq:A_longitude} with the gauge potentials of $\alpha$ and
$\beta$ monopoles (see Eq.(34), ($34^{\prime}\alpha$), ($34^{\prime}\beta$)
in Yang's paper\cite{yangGeneralizationDiracMonopole1978}), we can
conclude that $T^{2}=-1$ conical intersection behaves as the $\alpha$-type
Yang monopole when viewed from the upper state $(\Psi^{+})$, and
behaves as the $\beta$-type when viewed from the lower state ($\Psi^{-}$).

\subsection{Visualize Berry connection through Hopf fibration\label{subsec:Visualize-Berry-connection}}

Eqs. \ref{eq:A_R}, \ref{eq:A_theta} and \ref{eq:A_longitude} show
that in the current choice of gauge, Berry connection is non-zero
only along the longitude degree of freedom that is defined on the
$S^{3}$ residing within the subspace spanned by $\vec{e}_{x},\vec{e}_{y},\vec{e}_{u},\vec{e}_{v}$.
The property of Berry connection in the $S^{3}$-longitude space is
determined by $D_{\lambda}$ and $B_{\lambda}$ in Eq.\ref{eq:A_factor},
which will be called ``Berry connection factor'' in the following
discussions. The explicit mathematical formula for $D_{\lambda}$
and $B_{\lambda}$ in stereographic projection coordinates are already
derived in Yang's paper. However, because stereographic projection
is not metric-preserving and distorts the length, the geometric properties
of Berry connection factors are not obvious from these expressions.
In order to provide an intuitive understanding about their geometric
features, in Section \ref{subsec:constant_BD} we will first show
that the two Berry connection factors are constant along any great
circle within the $S^{3}$-longitude space. We will then discuss how
this property can be used to visualize the Berry connection factors
in Section \ref{subsec:Hopf}. All the $S^{3}$ mentioned in Section
\ref{subsec:constant_BD} and \ref{subsec:Hopf} refer to the $S^{3}$-longitude
space unless otherwise specified.

\subsubsection{Berry connection factors along a great circle are determined by its
bivector\label{subsec:constant_BD}}

The Berry connection factors in Eq.\ref{eq:A_factor} are defined
from the transition function $g_{l}$ in Eq.\ref{eq:g_transition}.
To connect with geometric algebra, we will first show how $g_{l}$
at point $l\in S^{3}$ is related to matrix representation of the
corresponding vector $\vec{l}$. As shown in Figure \ref{fig:coors}(b2),
the vector $\vec{l}$ can be written as $\vec{l}=\cos\phi\vec{e}_{x}+\sin\phi\vec{n}$.
Using the matrix representation of $\vec{e}_{x}$ in Eq.\ref{eq:Cl(5)}
and $\vec{n}$ in Eq.\ref{eq:vec_n}, $\vec{l}$ is related to the
transition function $g_{l}$ as
\begin{equation}
\vec{l}=\cos\phi\vec{e}_{x}+\sin\phi\vec{n}=\left(\begin{array}{cc}
 & e^{-\hat{n}\phi}\\
e^{\hat{n}\phi}
\end{array}\right)=\left(\begin{array}{cc}
 & g_{l}^{-1}\\
g_{l}
\end{array}\right)\label{eq:vec_l}
\end{equation}
This relation $\vec{l}=\left(\begin{array}{cc}
 & g_{l}^{-1}\\
g_{l}
\end{array}\right)$ provides a way to identify the transition function even when the
coordinates $(\phi,\mathbf{n}$) are not explicitly known.

The green circle illustrated in Figure \ref{fig:coors}(b2) represents
the type of great circle that passes through $\vec{e}_{x}$. Such
great circle is formed by the intersection of plane $\check{\beta}_{xn}$
(Eq.\ref{eq:bivector_xn}) with $S^{3}$ and will be denoted as $S_{xn}^{1}$.
We will also call $\check{\beta}_{xn}$ the bivector of $S_{xn}^{1}$.
The tangent direction along $S_{xn}^{1}$ (called ``along $S_{xn}^{1}$''
for short) is then represented by $\frac{\partial}{\partial\phi}$.
By substituting $g_{l}=e^{\hat{n}\phi}$ from Eq.\ref{eq:vec_l} into
Eq.\ref{eq:A_factor}, both Berry connection factors along $S_{xn}^{1}$
have constant values as

\begin{equation}
D_{\phi}(S_{xn}^{1})=\hat{n},\text{\text{ }}B_{\phi}(S_{xn}^{1})=-\hat{n}\label{eq:BD_xn}
\end{equation}
Note that because Berry connection factors take value in the Lie algebra
$\mathfrak{su}(2)$, the quaternion values in Eq.\ref{eq:BD_xn} should
be interpreted as $\mathfrak{su}(2)$ using Eq.\ref{eq:n_to_lie}.
Comparing Eq.\ref{eq:BD_xn} with $\check{\beta}_{xn}=\left(\begin{array}{cc}
\hat{n}\\
 & -\hat{n}
\end{array}\right)$ in Eq.\ref{eq:bivector_xn}, the Berry connection factors along $S_{xn}^{1}$
are related to its bivector as
\begin{equation}
\check{\beta}_{xn}=\left(\begin{array}{cc}
D_{\phi}(S_{xn}^{1})\\
 & B_{\phi}(S_{xn}^{1})
\end{array}\right)\label{eq:beta_xn}
\end{equation}
As we will prove below, the relation in Eq.\ref{eq:bivector_xn} is
generally true: for any great circle within the $S^{3}$-longitude
space, its bivector determines its Berry connection factors. In addition,
Eq.\ref{eq:BD_xn} also has the property that $B_{\phi}=-D_{\phi}$,
and we will show its generalization is that $B_{\phi}$ and $-D_{\phi}$
are related through adjoint transformation.

\begin{figure}[H]
	\begin{center}
		\includegraphics[width=2.75in]{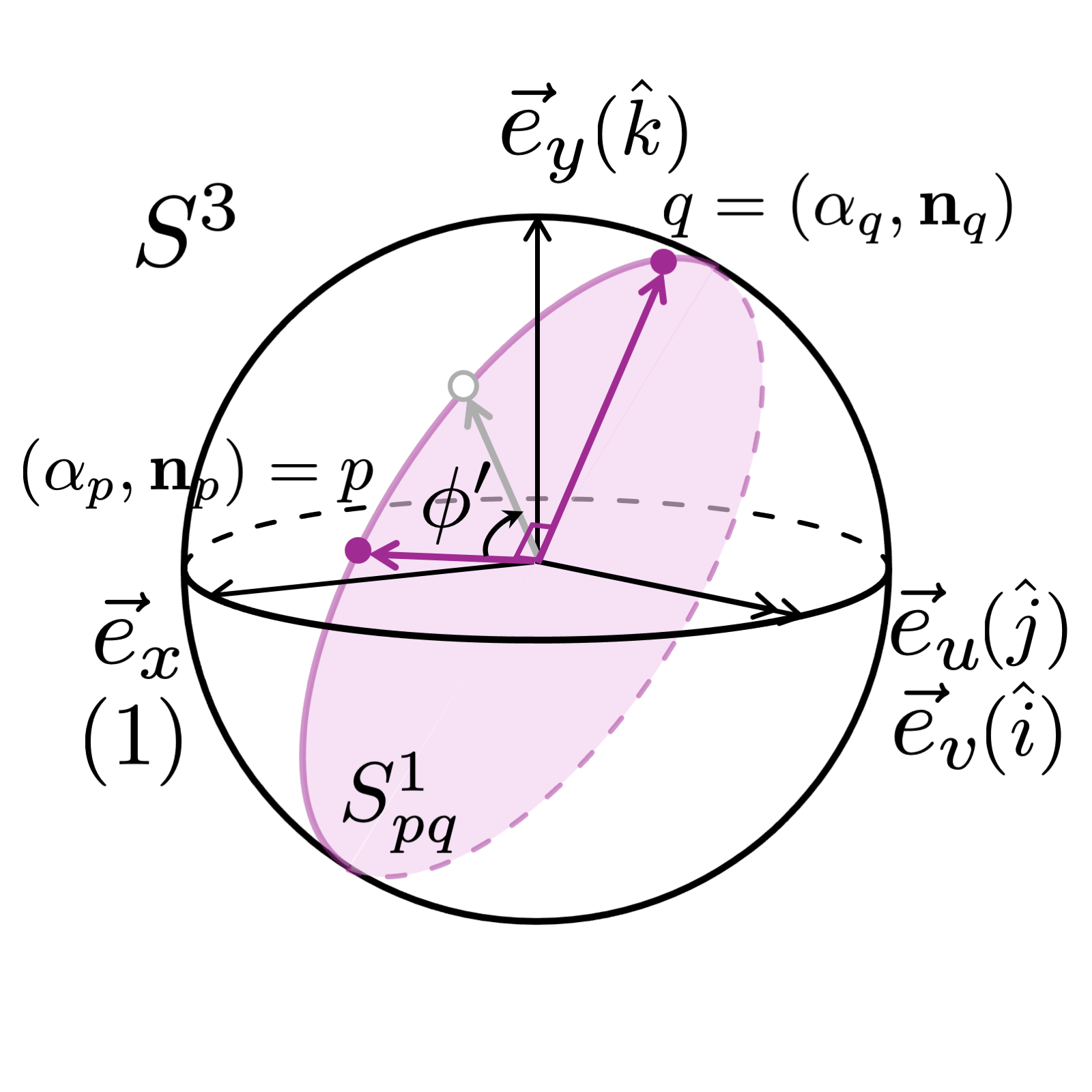}
		\caption{\label{fig:S1inS3}
Parametrization of a great circle (purple) on $S^{3}$ by
identifying two points $p,q$ on the great circle such that the two
unit vectors are orthogonal to each other. Coordinates $(\alpha_{p},\mathbf{n}_{p})$
for point $p$ follow Figure 1 (b2)-(b3), and similar for $q$. The
great circle is denoted as $S_{pq}^{1}$ correspondingly. This figure
also shows how a point on $S^{3}$ can be represented as a unit quaternion
by identifying the axes $\{\vec{e}_{x},\vec{e}_{y},\vec{e}_{u},\vec{e}_{v}\}$
with $\{1,\hat{k},\hat{j},\hat{i}\}$ respectively, which is used in Section \ref{subsec:Hopf}.
		}
		\end{center}
\end{figure}

Figure \ref{fig:S1inS3} illustrates an arbitrary great circle in
$S^{3}$ (purple), and we will now use geometric algebra to prove
that its Berry connection factors can be obtained from its bivector.
To compute the bivector, choose an arbitrary pair of orthogonal unit
vector on the circle, e.g. $p=(\alpha_{p},\mathbf{n}_{p})$ and $q=(\alpha_{q},\mathbf{n}_{q})$
such that $\vec{p}\perp\vec{q}$, and the great circle will be denoted
as $S_{pq}^{1}$ correspondingly. Using the matrix representation
of vector in Eq.\ref{eq:vec_l} and the definition of bivector in
Eq.\ref{eq:bivector}, the bivector of $S_{pq}^{1}$ is
\begin{equation}
\check{\beta}_{pq}=\vec{p}\vec{q}=\left(\begin{array}{cc}
e^{-\hat{n}_{p}\alpha_{p}}e^{\hat{n}_{q}\alpha_{q}}\\
 & e^{\hat{n}_{p}\alpha_{p}}e^{-\hat{n}_{q}\alpha_{q}}
\end{array}\right)\label{eq:beta_pq_1}
\end{equation}
Eq.\ref{eq:beta_pq_1} shows that bivector $\check{\beta}_{pq}$ takes
the diagonal form of
\begin{equation}
\check{\beta}_{pq}=\left(\begin{array}{cc}
\hat{b}_{+}\\
 & \hat{b}_{-}
\end{array}\right)\label{eq:beta_pq}
\end{equation}
Because bivector is a property of the plane, the values of $\hat{b}_{+}$
and $\hat{b}_{-}$ are independent of the choice of $p,q$. In addition,
$\vec{p}\perp\vec{q}$ implies $\vec{p}\vec{q}=-\vec{q}\vec{p}$ (see
Eq.\ref{eq:bivector}), leading to another way to compute $\check{\beta}_{pq}$
as
\begin{equation}
\check{\beta}_{pq}=-\vec{q}\vec{p}=\left(\begin{array}{cc}
-e^{-\hat{n}_{q}\alpha_{q}}e^{\hat{n}_{p}\alpha_{p}}\\
 & -e^{\hat{n}_{q}\alpha_{q}}e^{-\hat{n}_{p}\alpha_{p}}
\end{array}\right)\label{eq:beta_pq_2}
\end{equation}
Combining Eq.\ref{eq:beta_pq_1} and \ref{eq:beta_pq_2}, we get
\begin{equation}
\left\{ \begin{array}{c}
\text{\ensuremath{\hat{b}_{+}=}}e^{-\hat{n}_{p}\alpha_{p}}e^{\hat{n}_{q}\alpha_{q}}=-e^{-\hat{n}_{q}\alpha_{q}}e^{\hat{n}_{p}\alpha_{p}}\\
\hat{b}_{-}=e^{\hat{n}_{p}\alpha_{p}}e^{-\hat{n}_{q}\alpha_{q}}=-e^{\hat{n}_{q}\alpha_{q}}e^{-\hat{n}_{p}\alpha_{p}}
\end{array}\right.\label{eq:b1_b2}
\end{equation}
Eq.\ref{eq:b1_b2} shows that the quaternion numbers $\hat{b}_{+}$
and $\hat{b}_{-}$ satisfy the property $-\hat{b}_{i}=\hat{b}_{i}^{-1}$,
thus both must be pure unit quaternions.

Using matrix representations of $\vec{p}$ and $\vec{q}$ from Eq.\ref{eq:vec_l},
any point $S_{pq}^{1}$ can now be parametrized with the rotation
angle $\phi^{\prime}$ from $\vec{p}$ (see Figure \ref{fig:S1inS3})
as
\begin{equation}
\begin{array}{l}
S_{pq}^{1}\left(\phi^{\prime}\right)=\cos\phi^{\prime}\vec{p}+\sin\phi^{\prime}\vec{q} \\
=\left(\begin{array}{cc}
 & \cos\phi^{\prime}e^{-\hat{n}_{p}\alpha_{p}}+\sin\phi^{\prime}e^{-\hat{n}_{q}\alpha_{q}}\\
\cos\phi^{\prime}e^{\hat{n}_{p}\alpha_{p}}+\sin\phi^{\prime}e^{\hat{n}_{q}\alpha_{q}}
\end{array}\right)
\end{array}\label{eq:S1_pq}
\end{equation}
Based on the discussions about Eq.\ref{eq:vec_l}, the transition
function $g_{pq}(\phi^{\prime})$ along $S_{pq}^{1}$ can be identified
from the above matrix as
\begin{equation}
\left\{ \begin{array}{c}
g_{pq}^{-1}(\phi^{\prime})=\cos\phi^{\prime}e^{-\hat{n}_{p}\alpha_{p}}+\sin\phi^{\prime}e^{-\hat{n}_{q}\alpha_{q}}\\
g_{pq}(\phi^{\prime})=\cos\phi^{\prime}e^{\hat{n}_{p}\alpha_{p}}+\sin\phi^{\prime}e^{\hat{n}_{q}\alpha_{q}}
\end{array}\right.\label{eq:g_pq}
\end{equation}
One can show that $g_{pq}(\phi^{\prime})$ and $g_{pq}^{-1}(\phi^{\prime})$
indeed satisfy $g_{pq}(\phi^{\prime})g_{pq}^{-1}(\phi^{\prime})=1$
using Eq.\ref{eq:b1_b2}. By substituting Eq.\ref{eq:g_pq} into the
definition of Berry connection factors Eq.\ref{eq:A_factor}, $D_{\phi^{\prime}}$
and $B_{\phi^{\prime}}$ along $S_{pq}^{1}$ can be then computed
as
\begin{equation}
\begin{array}{c}
D_{\phi^{\prime}}(S_{pq}^{1})=-\frac{\partial g_{pq}^{-1}(\phi^{\prime})}{\partial\phi^{\prime}}g_{pq}(\phi^{\prime})\\
=\sin^{2}\phi^{\prime}e^{-\hat{n}_{p}\alpha_{p}}e^{\hat{n}_{q}\alpha_{q}}-\cos^{2}\phi^{\prime}e^{-\hat{n}_{q}\alpha_{q}}e^{\hat{n}_{p}\alpha_{p}}=\hat{b}_{+}
\end{array}
\end{equation}
\begin{equation}
\begin{array}{c}
B_{\phi^{\prime}}(S_{pq}^{1})=-\frac{\partial g_{pq}(\phi^{\prime})}{\partial\phi^{\prime}}g_{pq}^{-1}(\phi^{\prime})\\
=\sin^{2}\phi^{\prime}e^{\hat{n}_{p}\alpha_{p}}e^{-\hat{n}_{q}\alpha_{q}}-\cos^{2}\phi^{\prime}e^{\hat{n}_{q}\alpha_{q}}e^{-\hat{n}_{p}\alpha_{p}}=\hat{b}_{-}
\end{array}
\end{equation}
where we have used the two equivalent forms of $\hat{b}_{+}$ and
$\hat{b}_{-}$ in Eq.\ref{eq:b1_b2}. By comparing with $\check{\beta}_{pq}$
in Eq.\ref{eq:beta_pq},
\begin{equation}
\check{\beta}_{pq}=\left(\begin{array}{cc}
D_{\phi^{\prime}}(S_{pq}^{1})\\
 & B_{\phi^{\prime}}(S_{pq}^{1})
\end{array}\right)=\left(\begin{array}{cc}
e^{-\hat{n}_{p}\alpha_{p}}e^{\hat{n}_{q}\alpha_{q}}\\
 & -e^{\hat{n}_{q}\alpha_{q}}e^{-\hat{n}_{p}\alpha_{p}}
\end{array}\right)\label{eq:berry_pq}
\end{equation}
Eq.\ref{eq:berry_pq} proves that the bivector of the great circle
determines the Berry connection factors along its tangent direction.
Furthermore, Eq.\ref{eq:berry_pq} also shows that
\begin{equation}
B_{\phi^{\prime}}(S_{pq}^{1})=e^{\hat{n}_{p}\alpha_{p}}\left(-D_{\phi^{\prime}}(S_{pq}^{1})\right)e^{-\hat{n}_{p}\alpha_{p}}\label{eq:B_negD}
\end{equation}
This implies that $B$ and $-D$ along the same great circle are related
through an adjoint transformation. Therefore, Eq.\ref{eq:berry_pq}
and Eq.\ref{eq:B_negD} together conclude our proof of the two statements
proposed at the beginning of this section.

\subsubsection{Hopf fibration induced by Berry connection factors\label{subsec:Hopf}}

Eq.\ref{eq:berry_pq} shows how to compute the Berry connection factors
along any great circle. The factors $B$ and $D$ are formally 1-forms
that map a vector $\vec{v}_{p}\in T_{p}S^{3}$ (i.e. the tangent space
of $S^{3}$ at point $p$) to $\mathfrak{su}(2)$. From differential
geometry, the initial position $p$ and the initial velocity $\vec{v}_{p}\in T_{p}S^{3}$
uniquely determine a geodesic, which is a great circle in the case
of $S^{3}$. Therefore, Eq.\ref{eq:berry_pq} in fact fully specifies
the Berry connection factors at every point in $S^{3}$. In addition,
since $S^{3}\simeq SU(2)$, their corresponding tangent spaces must
also be isomorphic, i.e. $T_{p}S^{3}\simeq\mathfrak{su}(2)$. This
isomorphism implies that through Berry connection factor, every great
circle $S^{1}$ passing through a point $p$ is one-to-one mapped
to a Lie algebra $\mathfrak{g}\in\mathfrak{su}(2)$. Based on this
isomorphism, we can study the structure of Berry connections by examining
all the great circles in $S^{3}$ that are mapped to the same Lie
algebra value.

First, consider the Berry connection factor $D$ and denote the target
Lie algebra value as $\mathfrak{m}\in\mathfrak{su}(2)$, which translates
to quaternion $\hat{m}$ by Eq.\ref{eq:n_to_lie}. So far we have
been associating a point $(\phi,\mathbf{n})$ on $S^{3}$ with its
geometric algebra representation in Eq.\ref{eq:vec_l}, which consists
of $g_{l}=e^{\hat{n}\phi}$ and $g_{l}^{-1}=e^{-\hat{n}\phi}$ that
are inverse of each other. This implies that the transition function
$g_{l}=e^{\hat{n}\phi}$ itself can be used to specify the position
of a point on $S^{3}$ by identifying the axes $(\vec{e}_{x},\vec{e}_{v},\vec{e}_{u},\vec{e}_{y})$
with $(1,\hat{i},\hat{j},\text{\ensuremath{\hat{k}}})$ in quaternion
numbers respectively (shown in Figure \ref{fig:S1inS3}). By applying
this idea to Eq.\ref{eq:S1_pq}, points on a great circle (e.g. the
grey dot in Figure \ref{fig:S1inS3} ) can be parametrized with $\phi^{\prime}$
as
\begin{equation}
S_{pq}^{1}(\phi^{\prime})=\cos\phi^{\prime}e^{\hat{n}_{p}\alpha_{p}}+\sin\phi^{\prime}e^{\hat{n}_{q}\alpha_{q}}\label{eq:S1_pq_short}
\end{equation}
From Eq.\ref{eq:berry_pq} , in order for $S_{pq}^{1}$ to have the
desired factor $D(S_{pq}^{1})=\mathfrak{m}$, it needs to satisfy
\begin{equation}
e^{-\hat{n}_{p}\alpha_{p}}e^{\hat{n}_{q}\alpha_{q}}=\hat{m}\label{eq:D_condition}
\end{equation}
which can be rearranged as $e^{\hat{n}_{q}\alpha_{q}}=e^{\hat{n}_{p}\alpha_{p}}\hat{m}$.
By substituting this into Eq.\ref{eq:S1_pq_short}, a great circle
with Berry connection factor $D(S^{1})=\mathfrak{m}$ takes the form
\begin{equation}
S_{D}^{1}[\mathfrak{m}](\phi^{\prime})=e^{\hat{n}_{p}\alpha_{p}}e^{\hat{m}\phi^{\prime}}\label{eq:S1_D}
\end{equation}

Eq.\ref{eq:S1_D} in fact generates Hopf fibration $S^{1}\hookrightarrow S^{3}\to S^{2}$,
which decomposes $S^{3}$ as $S^{1}$-fibers on $S^{2}$-base. Some
features of the Hopf fibration include\cite{lyonsElementaryIntroductionHopf2003}
(1) there is a unique $S^{1}$ passing through every point in $S^{3}$,
and (2) every $S^{1}$ passes through a unique pair of antipodal points
on $S^{2}$. The great circles defined by Eq.\ref{eq:S1_D} satisfy
both properties. Property (1) is guaranteed by the isomorphism $T_{p}S^{3}\simeq\mathfrak{su}(2)$
discussed at the beginning of this section. For property (2), we can
choose the $S^{2}$-base as the one residing within the subspace $\{\vec{e}_{y},\vec{e}_{u},\vec{e}_{v}\}$,
and are identified with pure unit quaternions. One can then show that
at $\phi^{\prime}=\arctan\left(\frac{\cot\alpha_{p}}{\vec{n}_{p}\cdot\vec{m}}\right)$
and $\phi^{\prime}+\pi$, the real part of $e^{\hat{n}_{p}\alpha_{p}}e^{\hat{n}_{i}\phi^{\prime}}$
in Eq.\ref{eq:S1_D} vanishes, corresponding to the two antipodal
points where $S_{D}^{1}$ intersects with the $S^{2}$-base, thus
satisfying property (2). These considerations show that a given Lie
algebra value $\mathfrak{m}\in\mathfrak{su}(2)$ induces a Hopf fibration
through Eq.\ref{eq:S1_D} such that every $S^{1}$-fiber has Berry
connection factor $D$ equal to $\mathfrak{m}$ along its tangent
directions.

\begin{figure}[H]
	\begin{center}
		\includegraphics[width=3.5in]{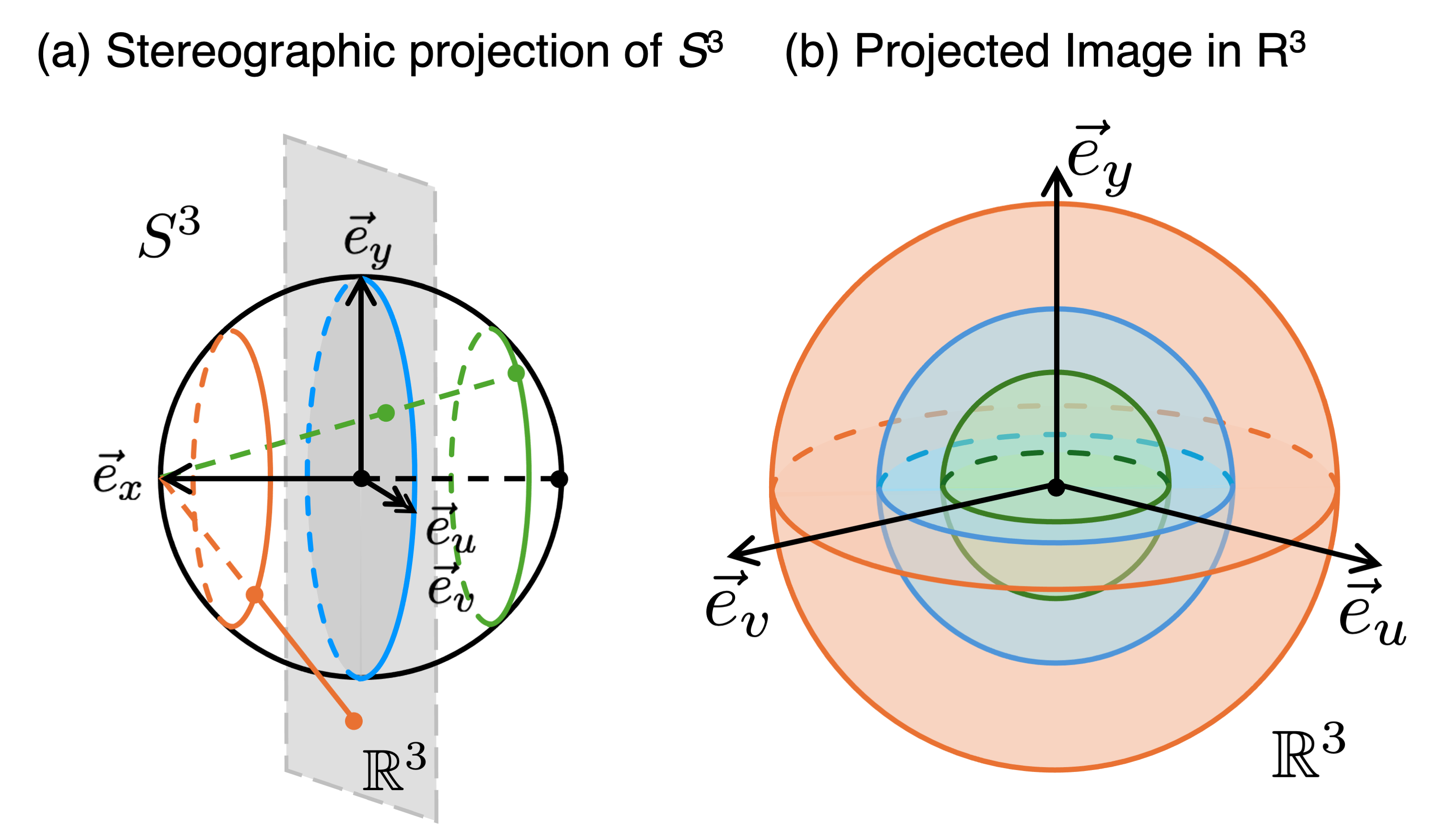}
		\caption{\label{fig:stereo}
Setup of stereographic projection of $S^{3}$ used in this
work for visualizing Berry connection factors. (a) illustrates how
points on $S^{3}$ are mapped to the image space $\mathbb{R}^{3}$.
The projection is applied from the tip of the unit vector $\vec{e}_{x}$.
The image space $\mathbb{R}^{3}$ is spanned by $\vec{e}_{y},\vec{e}_{u},\vec{e}_{v}$,
and is placed at the origin. (b) shows that the three colored $S^{2}$
in (a) are projected to the three colored spheres centered at the
origin in $\mathbb{R}^{3}$. The radius of the image sphere depends
on where the center of the original $S^{2}$ locates on the $x$ axis
in (a).
		}
		\end{center}
\end{figure}

\begin{figure}[H]
	\begin{center}
		\includegraphics[width=3.5 in]{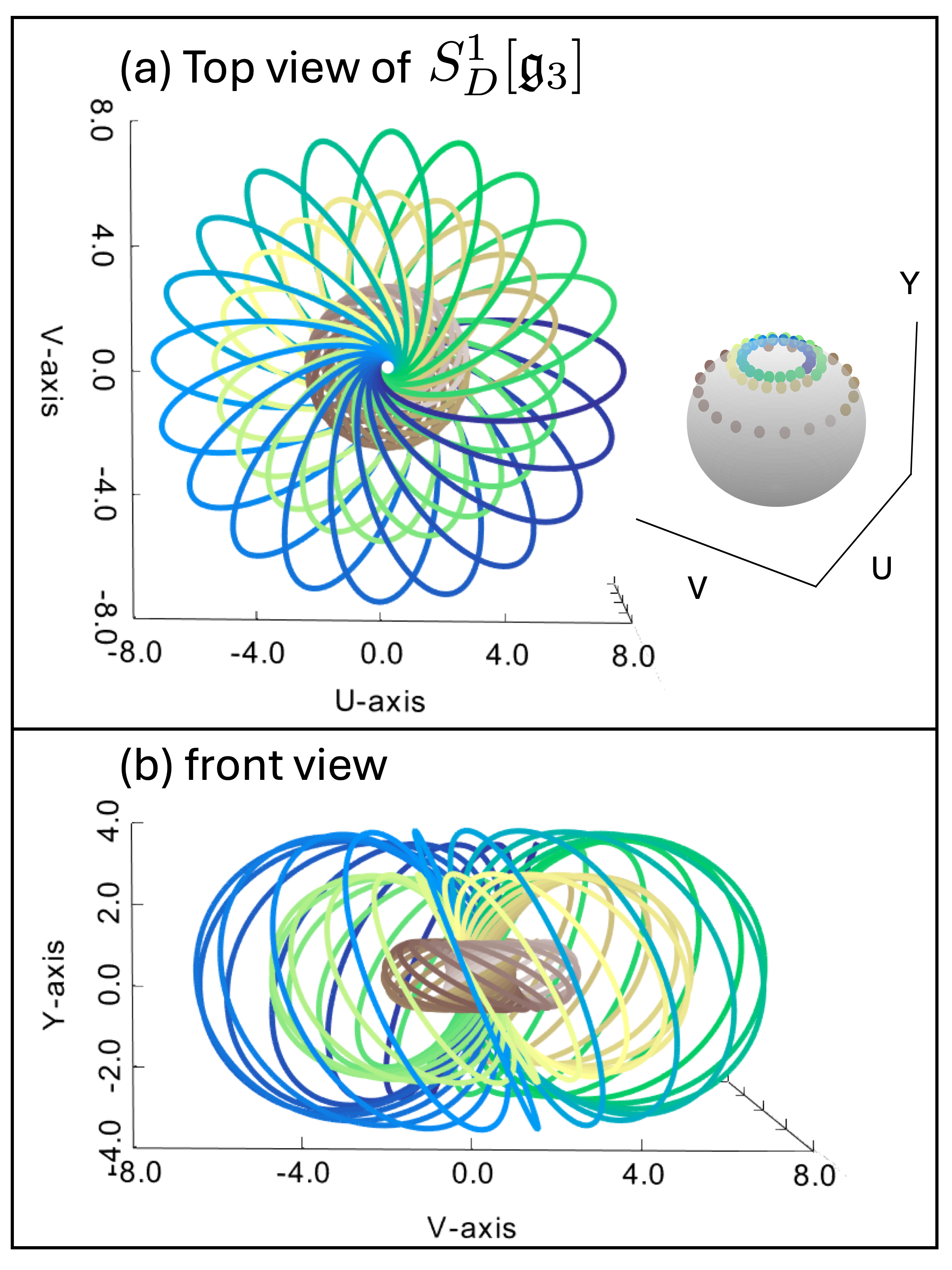}
		\caption{\label{fig:1donut}
Hopf fibration generated by Berry connection $S_{D}^{1}[\mathfrak{g}_{3}](p,\phi)$,
which is visualized by the stereographic projection in Figure \ref{fig:stereo}.
The inset of (a) shows the location of the initial points $p$ on
the $S^{2}$-base. The color of each initial point is consistent with
the color of the corresponding fiber.
		}
		\end{center}
\end{figure}

To visualize the Hopf fibration structure of Eq.\ref{eq:S1_D}, we
follow the work of Lyons,\cite{lyonsElementaryIntroductionHopf2003}
which visualizes the fibers in $\mathbb{R}^{3}$ through stereographic
projection. In this work, we apply the projection from the tip of
$\vec{e}_{x}$, and place the image plane $\mathbb{R}^{3}$ at the
origin in the subspace spanned by $\{\vec{e}_{y},\vec{e}_{u},\vec{e}_{v}\}$
(see Figure \ref{fig:stereo}(a)). Under this setup, every $S^{2}$
in $S^{3}$ that is perpendicular to $\vec{e}_{x}$ gets projected
to a $S^{2}$ centered at the origin in $\mathbb{R}^{3}$, and the
radius of the image sphere is determined by where the original $S^{2}$
locates on $\vec{e}_{x}$. As shown in Figure \ref{fig:stereo}(b),
the $S^{2}$-base (blue) maintains its size, those located on the
positive side of $\vec{e}_{x}$ gets magnified (orange), while those
on the negative side get reduced (green). As an example, we choose
$\mathfrak{g}_{3}$ as the target Lie algebra, and plot the Hopf-fibration
generated by $S_{D}^{1}[\mathfrak{g}_{3}]$ in Figure \ref{fig:1donut}.
As indicated by the colored dots in the inset of Figure \ref{fig:1donut},
the initial points we choose can be divided into three groups, each
group forming a ring perpendicular to $\vec{e}_{y}$ on the $S^{2}$-base.
The fibers generated from each ring weave into a torus that is perpendicular
to $\vec{e}_{y}$ in $\mathbb{R}^{3}$, thus we see three tori nested
together in Figure \ref{fig:1donut}. This nested tori structure is
the well-known characteristic of Hopf fibration.

\begin{figure}[H]
	\begin{center}
		\includegraphics[width=3.5in]{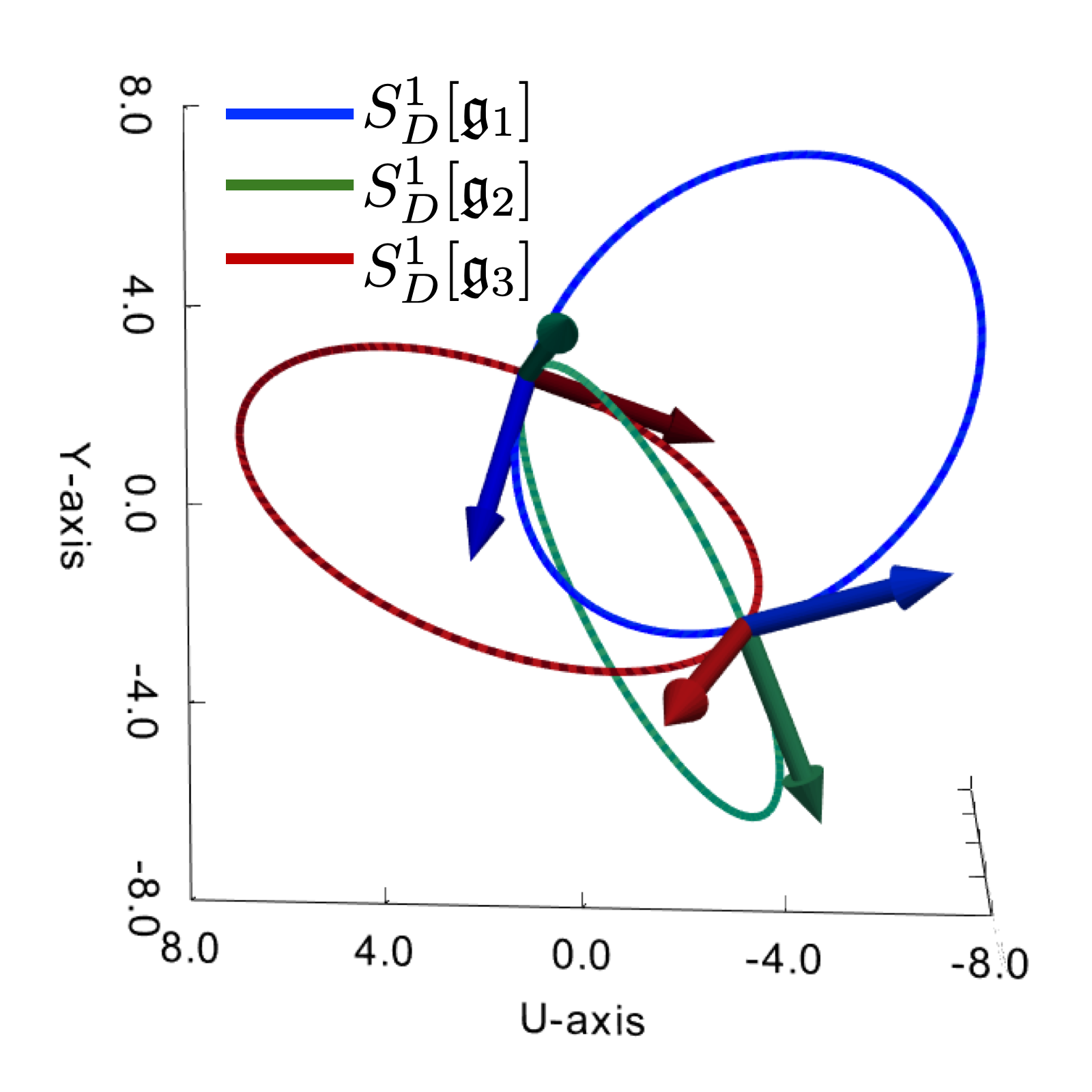}
		\caption{\label{fig:3ring}
Three $S^1$ fibers generated from $S_{D}^{1}[\mathfrak{g}_{i}]$,
$i=1,2,3$ respectively that intersect at the same two points. The
two intersecting points are $(0,-\frac{1}{\sqrt{3}},-\frac{1}{\sqrt{3}},-\frac{1}{\sqrt{3}})$
(lower right) and $(0,\frac{1}{\sqrt{3}},\frac{1}{\sqrt{3}},\frac{1}{\sqrt{3}})$
(upper left). The arrows indicate the positive tangent directions
(i.e. $+\mathfrak{g}_{i}$), and show that the three great circles
are mutually orthogonal at the intersecting points.
		}
		\end{center}
\end{figure}

\begin{figure}[H]
	\begin{center}
		\includegraphics[width=3.5in]{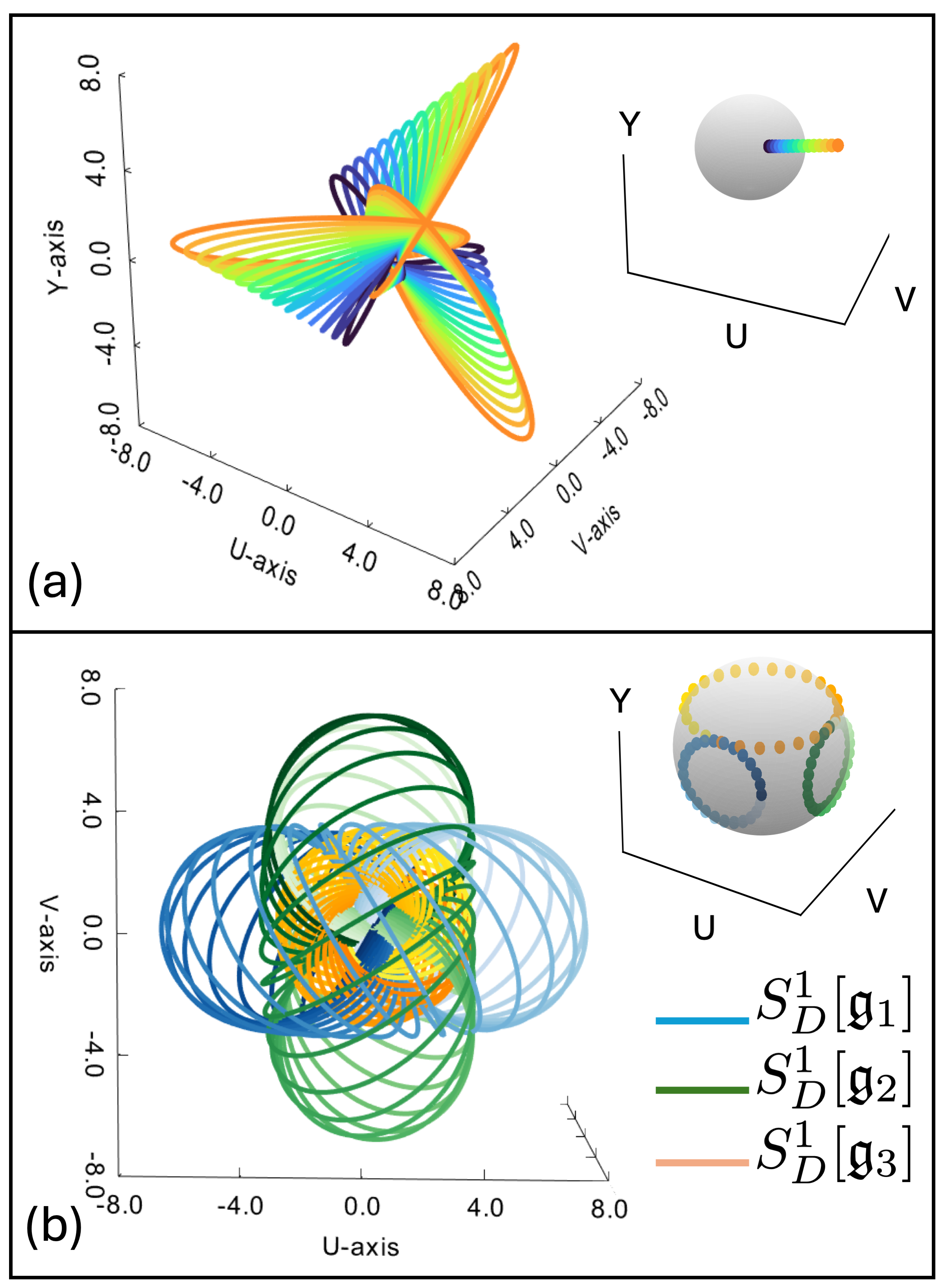}
		\caption{\label{fig:3donut}
Two ways of visualizing the Berry connection factor $D$
composed of three sets of Hopf-fibration $S_{D}^{1}[\mathfrak{g}_{i}]$,
$i=1,2,3$. (a) shows the three Hopf fibrations $S_{D}^{1}[\mathfrak{g}_{i}]$
generated from the same set of initial points that differ only in
$x$ coordinate, i.e. $p=\frac{1}{\sqrt{a^{2}+3}}(a+\hat{i}+\hat{j}+\hat{k}),a\in[0,2]$.
The initial points are shown as dots in the inset, and the three fibers
generated from the same initial point are colored by the same color
as the corresponding dot. (b) visualizes the three fibrations as three
mutually orthogonal tori. The initial points for $S_{D}^{1}[\mathfrak{g}_{i}]$
($i=1,2,3$) are selected from a ring perpendicular to the corresponding
axis (i.e. $\vec{e}_{v},\vec{e}_{u},\vec{e}_{y}$ respectively) on
the $S^{2}$-base, as shown in the inset.
		}
		\end{center}
\end{figure}

However, the visualization of Berry connection factor goes beyond
the standard Hopf-fibration structure in Figure \ref{fig:1donut}.
First, Berry connection factor is directional, i.e. there is a distinction
between positive and negative direction along each fiber corresponding
to $\pm\mathfrak{m}$. The positive directions are highlighted by
arrows in Figure \ref{fig:3ring}. In practice, we do not plot the
direction for each fiber, because once the positive direction of one
$S^{1}$-fiber is known, the rest can be determined by continuity.
Second, the Lie algebra $\mathfrak{su}(2)$ has three linearly independent
basis (e.g. $\mathfrak{g}_{1},\mathfrak{g}_{2},\mathfrak{g}_{3})$,
which correspondingly define three orthogonal tangent vectors at each
point $p$ that form the basis of $T_{p}S^{3}$ (i.e. the three arrows
in Figure \ref{fig:3ring}). As a result, the complete visualization
of Berry connection factors should be composed of three sets of Hopf-fibration,
i.e. $S_{D}^{1}[\mathfrak{g}_{1}]$, $S_{D}^{1}[\mathfrak{g}_{2}]$
and $S_{D}^{1}[\mathfrak{g}_{3}]$ . In particular, there exist three
$S^{1}$-fibers at every point $p\in S^{3}$, one from each Hopf-fibration,
intersecting orthogonally at that point. As an example, Figure \ref{fig:3ring}
shows the three fibers $S_{D}^{1}[\mathfrak{g}_{i}]$, $i,=1,2,3$
that are mutually orthogonal at their two intersecting points $p=e^{\hat{n}_{p}\alpha_{p}}=\hat{i}+\hat{j}+\hat{k}$
and $-\hat{i}-\hat{j}-\hat{k}$. In Figure \ref{fig:3donut}, we further
present two ways to visualize the three sets of Hopf-fibration. In
Figure \ref{fig:3donut}(a), we choose a set of initial points that
differ only in the $x$ coordinate, i.e. $p=e^{\hat{n}_{p}\alpha_{p}}=\frac{1}{\sqrt{a^{2}+3}}(a+\hat{i}+\hat{j}+\hat{k}),a\in[0,2]$.
The three sets of Hopf-fibration look like a ``fan'' with three
orthogonal blades, where each blade corresponds to one set of $S_{D}^{1}[\mathfrak{g}_{i}]$.
In Figure\ref{fig:3donut} (b), we choose initial points from three
rings on the $S^{2}$-base. The three rings are perpendicular to $\vec{e}_{v}$,
$\vec{e}_{u}$, $\vec{e}_{y}$ respectively, and are used to generate
fibers $S_{D}^{1}[\mathfrak{g}_{1}]$, $S_{D}^{1}[\mathfrak{g}_{2}]$,
$S_{D}^{1}[\mathfrak{g}_{3}]$ correspondingly. Similar to Figure
\ref{fig:1donut}, each group of fibers weaves into a torus that is
perpendicular to the corresponding axis. Together, the three sets
of Hopf fibration are presented as three donuts that are mutually
orthogonal.

\begin{figure}[H]
	\begin{center}
		\includegraphics[width=3in]{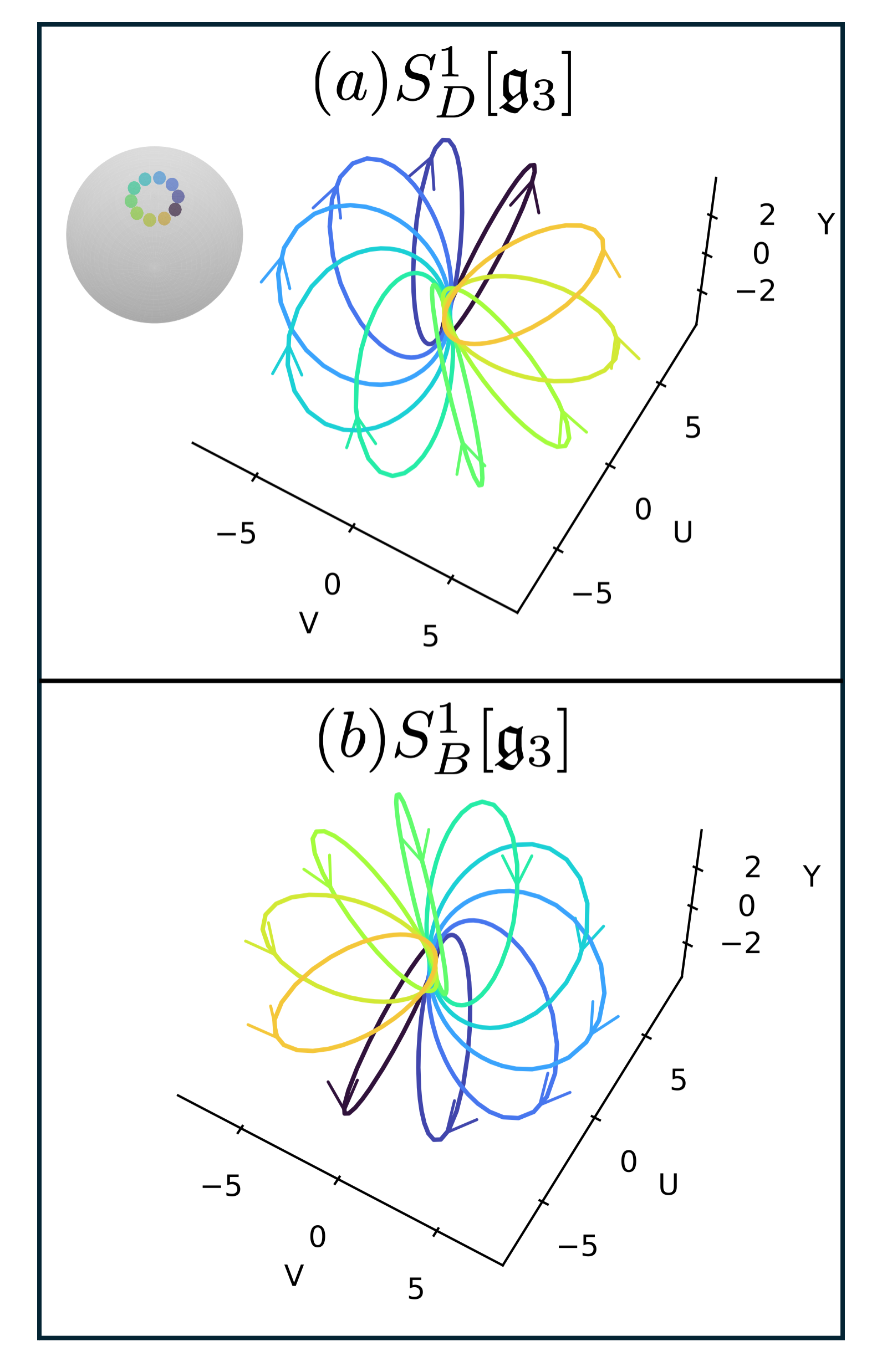}
		\caption{\label{fig:BvsD}
 Comparison between $S_{D}^{1}$ and $S_{B}^{1}$. In this
example, $\mathfrak{g}_{3}$ is used as the target Lie algebra. Both
are generated from the same set of initial points, shown in the inset
of (a). The arrow on each fiber indicates the tangent direction such that
Berry connection factor is positive, i.e. $+\mathfrak{g}_3$.
		}
		\end{center}
\end{figure}

The visualization of Berry connection factor $B$ can be generated
in similar ways. Given a target Lie algebra value $\mathfrak{m}\in\mathfrak{su}(2)$,
Eq.\ref{eq:berry_pq} shows that a great circle with $B(S^{1})=\mathfrak{m}$
should satisfy
\begin{equation}
-e^{\hat{n}_{q}\alpha_{q}}e^{-\hat{n}_{p}\alpha_{p}}=\hat{m}\label{eq:B_condition}
\end{equation}
which can be rearranged as $e^{\hat{n}_{q}\alpha_{q}}=-\hat{m}e^{\hat{n}_{p}\alpha_{p}}$.
By substituting this into Eq.\ref{eq:S1_pq_short}, such great circles
take the form of
\begin{equation}
S_{B}^{1}[\mathfrak{m}](\phi^{\prime})=e^{-\hat{m}\phi^{\prime}}e^{\hat{n}_{p}\alpha_{p}}\label{eq:S1_B}
\end{equation}
Figure \ref{fig:BvsD} compares the Hopf-fibration generated from
$S_{D}^{1}[\mathfrak{g}_{3}]$ and $S_{B}^{1}[\mathfrak{g}_{3}]$,
and the images are inversion from each other. To understand this relation,
consider $S_{D}^{1}[\mathfrak{m}]$ and $S_{B}^{1}[\mathfrak{m}]$
that pass through the same pair of antipodal points $\pm\hat{n}_{f}$
on the $S^{2}$-base. Let $e^{\hat{n}_{p}\alpha_{p}}=\hat{n}_{f}$
in Eq.\ref{eq:S1_D}, while $e^{\hat{n}_{p}\alpha_{p}}=-\hat{n}_{f}$
in Eq.\ref{eq:S1_B}. The corresponding fibers are parametrized as
$S_{D}^{1}[\mathfrak{m}](\phi^{\prime})=\hat{n}_{f}e^{\hat{m}\phi^{\prime}}$
and $S_{B}^{1}[\mathfrak{m}](\phi^{\prime})=e^{-\hat{m}\phi^{\prime}}(-\hat{n}_{f})$
respectively, which are related as $S_{B}^{1}[\mathfrak{m}](\phi^{\prime})\cdot S_{D}^{1}[\mathfrak{m}](\phi^{\prime})=1$,
i.e. inverse of each other. Note that two unit quaternions that are
inverse of each other have the same real part (mapped to $\vec{e}_{x}$)
but opposite imaginary part (mapped to $\vec{e}_{y},\vec{e}_{u},\vec{e}_{v}$).
Under the stereographic projection in Figure \ref{fig:stereo}(a),
their image points will be equal distance from the origin (determined
by the real part) and opposite in direction (determined by the imaginary
part), i.e. related through inversion.

\subsection{Berry curvature \label{subsec:Berry-curvature}}

The Lie algebra valued Berry curvature can be computed from Berry
connection as\cite{nakaharaGeometryTopologyPhysics2003}

\begin{equation}
F_{\mu\nu}=\partial_{\mu}A_{\nu}-\partial_{\nu}A_{\mu}+\left[A_{\mu},A_{\nu}\right]\label{eq:field_strength}
\end{equation}
Note that $F_{\mu\nu}$ in Eq. \ref{eq:field_strength} is related
to the field strength $f_{\mu\nu}$ defined in Yang's paper by a sign
of $-1$.\cite{yangGeneralizationDiracMonopole1978} From Eq.\ref{eq:A_R},
\ref{eq:A_theta} and \ref{eq:A_longitude}, $A_{R}=0$ and $\partial_{R}A_{\xi}=0$
($\xi$ denotes an arbitrary coordinate). Substitute into Eq.\ref{eq:field_strength},
we get
\begin{equation}
F_{R\xi}=-F_{\xi R}=0\label{eq:F_Rx}
\end{equation}
Eq.\ref{eq:F_Rx} suggests that non-zero Berry curvature only exists
on surface element perpendicular to the radial direction. In the case
of abelian U(1) group of $T^{2}=0$, Eq.\ref{eq:field_strength} is
further simplified as $F_{\mu\nu}=\partial_{\mu}A_{\nu}-\partial_{\nu}A_{\mu}$,
leading to the well-known result that its Berry curvature is uniform
on $S^{2}$ (see Table \ref{table:3Dvs5D}).

\begin{figure}[H]
	\begin{center}
		\includegraphics[width=3.5in]{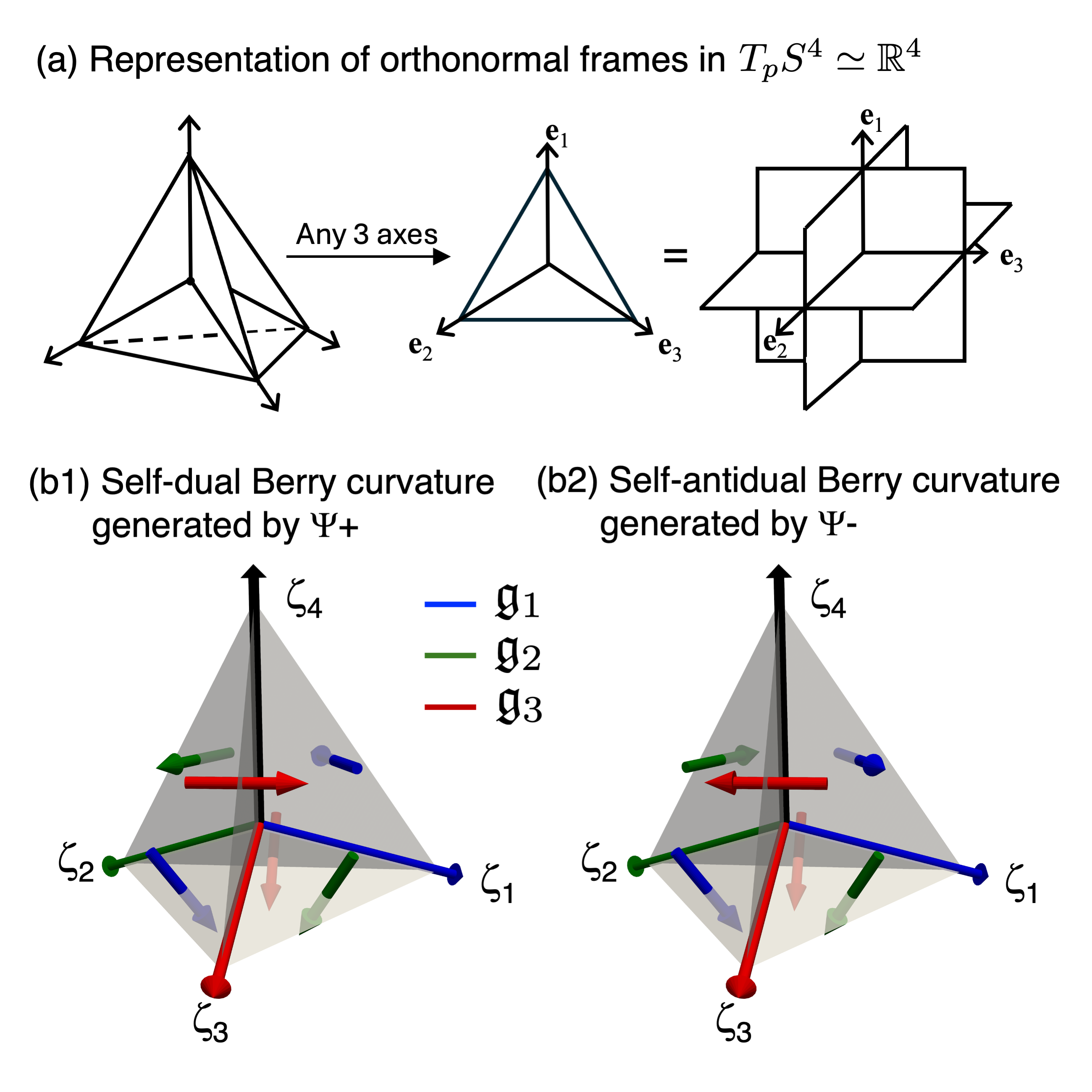}
		\caption{\label{fig:Fuv}
Illustration of the Berry curvature, which is defined on
$T_{p}S^{4}\simeq\mathbb{R}^{4}$ for every point $p\in S^{4}$. Because
there are 6 mutually orthogonal planes in $\mathbb{R}^{4}$, (a) shows
how the 6 planes are represented on a tetrahedron. (b1)-(b2) illustrate
the Berry curvature of $\Psi^{+}$and $\Psi^{-}$respectively. The
coordinates $\{\zeta_{i},1\leq i\leq4\}$ are defined in Eq.\ref{eq:coordinate}
in the main text, and the axes are colored by the Lie algebra value
$\mathfrak{g}_{i}$ of the Berry connection factor $D$ along that
direction (black means zero). The Berry curvature component $F_{\mu\nu}$
is represented by the arrow that goes through the plane spanned by
$\zeta_{u}$ and $\zeta_{v}$. Each arrow is colored by the Lie algebra
value $\mathfrak{g}_{i}$ of $F_{\mu\nu}$, and is pointing in the
positive direction (i.e. $+\mathfrak{g}_{i}$).
		}
		\end{center}
\end{figure}

For $T^{2}=-1$ conical intersection, Eq.\ref{eq:F_Rx} implies that
nonzero Berry curvature only exists on $S^{4}$ surface, and Figure
\ref{fig:Fuv}(a) illustrates how we can graphically represent all
surface elements in $T_{p}S^{4}\simeq\mathbb{R}^{4}$ using a tetrahedron.
As discussed at the end of Section \ref{subsec:Yang-monopole}, the
$T^{2}=-1$ conical intersection behaves as the $\alpha$ and $\beta$-type
Yang monopole when viewed from $\Psi^{+}$ and $\Psi^{-}$ respectively.
Yang has proved that given a choice of Lie algebra orthonormal basis
( e.g. $\{\mathfrak{g}_{1},\mathfrak{g}_{3},\mathfrak{g}_{3}\}$),
at every point $p\in S^{4}$, there exists an orthonormal coordinate
system $\{\zeta_{1},\zeta_{2},\zeta_{3},\zeta_{4}\}$ in the tangent
space $T_{p}S^{4}\simeq\mathbb{R}^{4}$, such that the field strength
$f_{\mu\nu}$ of the $\alpha$ and $\beta$-type monopoles takes a
standard form:
\begin{equation}
\begin{array}{cc}
F_{\mu\nu}=-f_{\mu\nu}=-a\left(\begin{array}{cccc}
0 & \mathfrak{g}_{3} & -\mathfrak{g}_{2} & c\cdot\mathfrak{g}_{1}\\
-\mathfrak{g}_{3} & 0 & \mathfrak{g}_{1} & c\cdot\mathfrak{g}_{2}\\
\mathfrak{g}_{2} & -\mathfrak{g}_{1} & 0 & c\cdot\mathfrak{g}_{3}\\
-c\cdot\mathfrak{g}_{1} & -c\cdot\mathfrak{g}_{2} & -c\cdot\mathfrak{g}_{3} & 0
\end{array}\right), \\
c=\left\{ \begin{array}{cc}
+1 & \alpha\text{-type}\\
-1 & \beta\text{-type}
\end{array}\right.
\end{array}\label{eq:F_standard}
\end{equation}
In Eq.\ref{eq:F_standard}, $c=\pm1$ implies the field is ``self-dual''
and ``self-antidual'' respectively, and $a$ is a constant describing
the ``strength of monopole''. Yang did not explicitly define what
the orthonormal coordinate system is, and we will show that the results
from Section \ref{subsec:Hopf} in fact provide a simple way to identify
such coordinates.

The expressions for field strength $f_{\mu\nu}$ have already been
derived by Yang, and here we mainly rewrite those equations in terms
of Berry curvature $F_{\mu\nu}$. Note that the Berry curvature is
antisymmetric, i.e. $F_{\mu\nu}=-F_{\nu\mu}$ (see Eq.\ref{eq:F_standard}).
Let $\theta$ and $\lambda$ be the latitude and longitude parameter
on $S^{4}$. Substitute $A_{\theta}=0$ (Eq.\ref{eq:A_theta}) into
Eq.\ref{eq:field_strength}, we get
\begin{equation}
F_{\lambda\theta}=-\partial_{\theta}A_{\lambda}\label{eq:F_thetax}
\end{equation}
Substitute $A_{\lambda}$ in Eq.\ref{eq:A_longitude} into Eq.\ref{eq:F_thetax},
$F_{\lambda\theta}$ are related to Berry connection factors as
\begin{equation}
\begin{array}{c}
F_{\lambda\theta}^{+,N}=-\frac{\sin\theta}{2}D_{\lambda},\text{ }F_{\lambda\theta}^{+,S}=\frac{\sin\theta}{2}B_{\lambda}\\
F_{\lambda\theta}^{-,N}=-\frac{\sin\theta}{2}B_{\lambda},\text{ }F_{\lambda\theta}^{-,S}=\frac{\sin\theta}{2}D_{\lambda}
\end{array}\label{eq:F_theta_long}
\end{equation}
In addition, using $\partial_{\mu}B_{\nu}-\partial_{\nu}B_{\mu}+\left[B_{\mu},B_{\nu}^ {}\right]=0$
for $\mu,\nu\in S^{3}$ (and same for $D$) proved in Yang's paper,
$F_{\lambda\lambda^{\prime}}$ in Eq.\ref{eq:field_strength} are
related to Berry connection factors as
\begin{equation}
\begin{array}{c}
F_{\lambda\lambda^{\prime}}^{+,N}=F_{\lambda\lambda^{\prime}}^{-,S}=-\frac{\sin^{2}\theta}{4}\left[D_{\lambda},D_{\lambda^{\prime}}\right]\\
F_{\lambda\lambda^{\prime}}^{+,S}=F_{\lambda\lambda^{\prime}}^{-,N}=-\frac{\sin^{2}\theta}{2}\left[B_{\text{\ensuremath{\lambda}}},B_{\lambda^{\prime}}\right]
\end{array}\label{eq:F_longlong}
\end{equation}
Eq.\ref{eq:F_theta_long} and Eq.\ref{eq:F_longlong} completely specify
the Berry curvature on $S^{4}$ surface.

To identify the desired orthonormal coordinate $\{\zeta_{1},\zeta_{2},\zeta_{3},\zeta_{4}\}$
that produces the standard form of Berry curvature, first consider
$F^{+,N}$ and $F^{-,S}$ that both depend on factor $D$. Because
the latitude is orthogonal to the longitude space, $\theta$ is naturally
chosen as one of the coordinates $\zeta_{4}$ ( the black axis in
Figure \ref{fig:Fuv}(b)). The remaining three coordinates $\zeta_{1,2,3}$
should thus all be in the $S^{3}$-longitude space. Using Eq.\ref{eq:S1_D},
at a particular point $p=(\alpha_{p},\mathbf{n}_{p})$ , the three
Lie algebra basis $\{\mathfrak{g}_{i}\}$ determine three great circles
$S_{D}^{1}[\mathfrak{g}_{i}]$, $i=\{1,2,3\}$ with angular parameters
$\phi_{i}$ respectively. By the definition of $S_{D}^{1}[\mathfrak{g}_{i}]$,
we have
\begin{equation}
\begin{array}{c}
D_{\phi_{i}}(p)=\mathfrak{g}_{i}\\
\left[D_{\phi_{i}}(p),D_{\phi_{j}}(p)\right]=[\mathfrak{g_{i},\mathfrak{g}_{j}}]=2\epsilon_{ijk}\mathfrak{g}_{k}
\end{array}\label{eq:D_ijk}
\end{equation}
From Section \ref{subsec:Hopf}, the three tangent vectors $\frac{\partial}{\partial\phi_{i}}$
are mutually orthogonal at $p$, thus determine the directions of
the remaining three coordinates. This is illustrated in Figure \ref{fig:Fuv}(b),
where the remaining three axes correspond to the three tangent vectors
in Figure \ref{fig:3ring}. Based on these considerations, our choice
of the orthonormal coordinate system is
\begin{equation}
\begin{array}{c}
d\zeta_{i}=\sin\theta d\phi_{i},i=1,2,3\\
d\zeta_{4}=d\theta
\end{array}\label{eq:coordinate}
\end{equation}
Substitute Eq.\ref{eq:D_ijk} and Eq.\ref{eq:coordinate} into Eq.\ref{eq:F_theta_long}
and Eq.\ref{eq:F_longlong}, the Berry curvature in the chosen coordinate
system become

\begin{equation}
\begin{array}{c}
F_{k4}^{+,N}=-\frac{1}{2}\mathfrak{g}_{k},\text{ }F_{k4}^{-,S}=\frac{1}{2}\mathfrak{g}_{k}\\
F_{ij}^{+,N}=F_{ij}^{-,S}=-\frac{1}{2}\epsilon_{ijk}\mathfrak{g}_{k}
\end{array}\label{eq:curvature}
\end{equation}
By comparing Eq.\ref{eq:curvature} with the standard form in Eq.\ref{eq:F_standard},
the Berry curvature of $\Psi^{+}$ is indeed in the standard self-dual
form (i.e. $\alpha$-type), while $\Psi^{-}$ is in the standard self-antidual
form (i.e. $\beta$-type). The Berry curvatures in Eq.\ref{eq:curvature}
are represented as arrows in Figure \ref{fig:Fuv}(b), where the arrow
is colored by the type of $\mathfrak{g}_{i}$, and is pointing in
the surface direction with $+\mathfrak{g}_{i}$. Furthermore, the
prefactor $\frac{1}{2}$ in Eq.\ref{eq:curvature} implies that $T^{2}=-1$
conical intersection behaves as Yang monopole with strength $|a|=\frac{1}{2}$,
leading to second Chern number $+\frac{1}{2^{2}}$ for $\Psi^{+}$
and $-\frac{1}{2^{2}}$ for $\Psi^{-}$. This is consistent with the
property of $T^{2}=0$ conical intersection that it behaves as a Dirac
monopole with strength $|g|=\frac{1}{2}$.

Lastly, it's worth mentioning that using the same coordinates in Eq.\ref{eq:coordinate},
it's possible to obtain similar standard form for $F^{-,N}$ and $F^{+,S}$
that both depend on factor $B$. Recall from Eq.\ref{eq:B_negD},
$B_{\phi}$ is related to $-D_{\phi}$ by adjoint transformation $B_{\phi_{i}}=e^{\hat{n}_{p}\alpha_{p}}(-D_{\phi_{i}})e^{-\hat{n}_{p}\alpha_{p}}$.
One important property of adjoint transformation is that it preserves
Lie bracket, leading to
\begin{equation}
\left[B_{\phi_{i}},B_{\phi_{j}}\right]=e^{\hat{n}_{p}\alpha_{p}}\left[D_{\phi_{i}},D_{\phi_{j}}\right]e^{-\hat{n}_{p}\alpha_{p}}
\end{equation}
where we have used the fact that $\left[D_{\phi_{i}},D_{\phi_{j}}\right]=\left[-D_{\phi_{i}},-D_{\phi_{j}}\right]$.
By defining new basis of Lie algebra, i.e. $\tilde{\mathfrak{g}}_{i}=e^{\hat{n}_{p}\alpha_{p}}\mathfrak{g}_{i}e^{-\hat{n}_{p}\alpha_{p}}$,$i=1,2,3$,
one can again show that $F^{-,N}$ and $F^{+,S}$ take the standard
form of Eq.\ref{eq:F_standard} in $\tilde{\mathfrak{g}}_{i}$.

\section{Conclusion and outlook}

In this work, we study the topological and geometric properties of
$T^{2}=-1$ conical intersections. The combination of geometric algebra
and quaternion numbers provides a simple way to derive its eigenfunctions,
Berry connection and Berry curvature. When viewed from the upper or
lower states, the $T^{2}=-1$ conical intersection behaves as the
self-dual or self-antidual Yang monopole respectively. In addition,
by showing that every Lie algebra value of the Berry connection factor
induces its own Hopf-fibration on a $S^{3}$ longitude space, we can
visualize its structure using stereographic projection.

Physicists often view molecules as 0D objects and don't consider their
topological properties to be particularly interesting.\cite{grushinIntroductionTopologicalPhases}
However, from the chemists' perspective, molecules are in fact high-dimensional
object because most interesting chemical phenomena involve exploring
the molecular configuration space, which can potentially give arise
to complicated topological and geometric structure. The $T^{2}=-1$
molecular conical intersection studied in this work is only an initial
step towards studying topological phenomena in chemistry, and there
are still many questions need to be answered in future studies.

As mentioned in the introduction, one unique property of molecular
conical intersections is its high dimensional degeneracy space, i.e.
the ``seam''.\cite{matsikaEffectsSpinorbitCoupling2001} Recently,
Morreale et.al have proved several topological properties of $T^{2}=+1$
seam, such as a crossing seam has no boundary, \cite{morrealeTopologyConicalIntersection2024}
and it would be interesting to see how these properties can be generalized
to $T^{2}=-1$ conical intersection. It is also well-known that the
geometric phase of $T^{2}=+1$ conical intersection is simply determined
by the ``winding'' number around the seam.\cite{vanderbiltBerryPhasesElectronic2018}
The analysis of geometric phase for $T^{2}=-1$ conical intersection
will likely be more complicated when taking the seam into account,
which may be affected by the curvature of the seam\cite{yarkonyConnectivitySeamsConical2005}
and require choosing a proper set of orthogonal nuclear displacements
to describe the branching space.\cite{hanPropertiesSeamBranching2003}

The conical intersections discussed in this work are two-level ones,
and a natural question is how the properties will differ when three-levels
or more are involved. The three-level $T^{2}=+1$ conical intersection
is relatively better understood both analytically and numerically.\cite{coeExtentConnectivityConical2008,matsikaThreeStateConicalIntersections2005,schuurmanCharacterizationThreeState2006}
Its model Hamiltonian is $3\times3$ real symmetric matrix, thus degeneracy
requires two diagonal conditions and three off-diagonal conditions,
resulting in a five-dimensional branching space. From Section \ref{subsec:modelH},
the model Hamiltonian of three-level $T^{2}=-1$ conical intersection
should be $3\times3$ quaternion Hermitian matrix. The degeneracy
requires two diagonal conditions in real numbers and three off-diagonal
conditions in quaternion numbers, resulting in a 14-dimensional branching
space. Yang has proved that there is no SU(2) monopole with SO(N)
symmetry for $N\geq6$.\cite{yangGeneralizationDiracMonopole1978}
This implies that the three-level $T^{2}=-1$ conical intersection
does not have full rotational symmetry in the branching space, i.e.
SO(14). As a result, the study of its properties may require a different
mathematical approach compared to the two-level ones.

In terms of numerical simulations, it will require the combination
of electronic structure theory, geometry optimization algorithm and
quantum dynamics method to better understand the role of $T^{2}=-1$
conical intersections in photochemistry, especially whether the non-abelian
geometric phase can affect the excited state dynamics in some unique
way apart from the better studied U(1) geometric phase. The visualization
results in this work can be used to verify whether the numerical Berry
connections are implemented correctly, similar to how correct numerical
implementation for $T^{2}=+1$ conical intersection should reproduce
the analytical behavior.\cite{hohensteinAnalyticFormulationDerivative2016}
Recently, there have been progresses towards non-adiabatic quantum
dynamics involving degenerate states.\cite{wuChemicalReactionRates2020,bianModelingNonadiabaticDynamics2021}
The analysis in this work suggests that it may worth considering developing
the corresponding quantum dynamics simulations directly in quaternion
numbers, given that it is the natural number system for $T^{2}=-1$
system.

\section*{Acknowledgements}

I would like to thank Prof.Wu-Yi Hsiang for teaching me differential
geometry and Lie group theory in his free time, which plant the seed
for this work. This work is supported by National Science Foundation
Award Number 2440880.

\section*{Supplementary material}

The supplementary material includes the python script that compute
and plot Figure \ref{fig:1donut} through Figure \ref{fig:BvsD} .

\section*{Author declaration}

The author has no conflicts to disclose.

\section*{Data availability}

All data has been provided either in the main text or in the supplementary
material.

\newpage
%
\end{document}